\documentclass{JINST}
\usepackage{graphicx}
\usepackage{xspace}
\usepackage{amsmath,amssymb,epsfig}

\newcommand{\GeVc}{\ensuremath{\mbox{GeV}/c}\xspace}
\newcommand{\MeVc}{\ensuremath{\mbox{MeV}/c}\xspace}

\newcommand{\cm}{\ensuremath{\mbox{cm}}\xspace}
\newcommand{\mm}{\ensuremath{\mbox{mm}}\xspace}

\newcommand{\micros}{\ensuremath{\mu \mbox{s}}\xspace}

\newcommand{\dedx}{\ensuremath{\mbox{d}E/\mbox{d}x}\xspace}
\newcommand{\pip}{\ensuremath{\pi^+}\xspace}
\newcommand{\pim}{\ensuremath{\pi^-}\xspace}
\newcommand{\piz}{\ensuremath{\pi^0}\xspace}

\newcommand{\dzeroprime}{\ensuremath{d'_0}\xspace}

\newcommand{\evtspill}{\ensuremath{N_{\mathrm{evt}}}\xspace}
\newcommand{\pt}{\ensuremath{p_{\mathrm{T}}}\xspace}

\def\be{\begin{equation}}
\def\ee{\end{equation}}
\def\bea{\begin{eqnarray}}
\def\eea{\end{eqnarray}}

\title{Absolute Momentum Calibration of the HARP TPC }

\author{%
M.G.~Catanesi, 
E.~Radicioni \\
Universit\`{a} degli Studi e Sezione INFN, Bari, Italy}
\author{R.~Edgecock, 
M.~Ellis$^{1}$,          
F.J.P.~Soler$^{2}$ \\
Rutherford Appleton Laboratory, Chilton, Didcot, UK}
\author{%
C.~G\"{o}\ss ling \\
Institut f\"{u}r Physik, Universit\"{a}t Dortmund, Germany} 
\author{%
S.~Bunyatov, 
A.~Krasnoperov, 
B.~Popov$^{3}$, 
V.~Serdiouk,        
V.~Tereschenko  \\
Joint Institute for Nuclear Research, JINR Dubna, Russia} 
\author{
E.~Di~Capua, 
G.~Vidal--Sitjes$^{4}$  \\ 
Universit\`{a} degli Studi e Sezione INFN, Ferrara, Italy}  
\author{
A.~Artamonov$^{5}$,   
S.~Giani, 
S.~Gilardoni,       
P.~Gorbunov$^{5}$,  
A.~Grant,  
A.~Grossheim$^{7}$, 
V.~Ivanchenko$^{8}$,  
A.~Kayis-Topaksu$^{9}$,
J.~Panman, 
I.~Papadopoulos,  
E.~Tcherniaev, 
I.~Tsukerman$^{5}$,   
R.~Veenhof, 
C.~Wiebusch$^{10}$,    
P.~Zucchelli$^{6,11}$ \\ 
CERN, Geneva, Switzerland} 
\author{
A.~Blondel, 
S.~Borghi$^{12}$,  
M.C.~Morone$^{13}$, 
G.~Prior$^{14}$,   
R.~Schroeter \\
Section de Physique, Universit\'{e} de Gen\`{e}ve, Switzerland} 
\author{
C.~Meurer \\
Institut f\"{u}r Physik, Forschungszentrum Karlsruhe, Germany} 
\author{
U.~Gastaldi  \\
Laboratori Nazionali di Legnaro dell' INFN, Legnaro, Italy} 
\author{
G.~B.~Mills$^{15}$ \\
Los Alamos National Laboratory, Los Alamos, USA} 
\author{
J.S.~Graulich$^{16}$, 
G.~Gr\'{e}goire  \\
Institut de Physique Nucl\'{e}aire, UCL, Louvain-la-Neuve,Belgium} 
\author{
M.~Bonesini, 
F.~Ferri       \\     
Sezione INFN Milano Bicocca, Universit\`{a} degli Studi Milano Bicocca, 
Milano, Italy} 
\author{
M.~Kirsanov \\
Institute for Nuclear Research, Moscow, Russia} 
\author{
A. Bagulya, 
V.~Grichine,  
N.~Polukhina  \\
P. N. Lebedev Institute of Physics (FIAN), Russian Academy of
Sciences, Moscow, Russia} 
\author{
V.~Palladino \\
Universit\`{a} ``Federico II'' e Sezione INFN, Napoli, Italy} 
\author{
L.~Coney$^{15}$, 
D.~Schmitz$^{15}$ \\
Columbia University, New York, USA}  %
\author{
G.~Barr, 
A.~De~Santo$^{17}$ \\ 
Nuclear and Astrophysics Laboratory, University of Oxford, UK} 
\author{
F.~Bobisut, 
D.~Gibin,
A.~Guglielmi, 
M.~Mezzetto
\thanks{Corresponding author, e-mail: Mauro.Mezzetto@pd.infn.it} \\
Universit\`{a} degli Studi e Sezione INFN, Padova, Italy} 
\author{
J.~Dumarchez \\
LPNHE, Universit\'{e}s de Paris VI et VII, Paris, France} 
\author{
U.~Dore \\
Universit\`{a} ``La Sapienza'' e Sezione INFN Roma I, Roma, Italy} 
\author{
D.~Orestano, 
F.~Pastore, 
A.~Tonazzo, 
L.~Tortora \\
Universit\`{a} degli Studi e Sezione INFN Roma III, Roma, Italy} 
\author{
C.~Booth, 
L.~Howlett \\
Dept. of Physics, University of Sheffield, UK} 
\author{
M.~Bogomilov, 
M.~Chizhov, 
D.~Kolev, 
R.~Tsenov \\
Faculty of Physics, St. Kliment Ohridski University, Sofia, Bulgaria} 
\author{
S.~Piperov, 
P.~Temnikov \\
Institute for Nuclear Research and Nuclear Energy, 
Academy of Sciences, Sofia, Bulgaria} 
\author{
M.~Apollonio, 
P.~Chimenti,  
G.~Giannini \\
Universit\`{a} degli Studi e Sezione INFN, Trieste, Italy} 
\author{
J.~Burguet--Castell, 
A.~Cervera--Villanueva, 
J.J.~G\'{o}mez--Cadenas, 
J. Mart\'{i}n--Albo,
P.~Novella, 
M.~Sorel \\
Instituto de F\'{i}sica Corpuscular, IFIC, CSIC and Universidad de Valencia \\
\vskip 30 pt
\llap{$^{~1}$}{Now at FNAL, Batavia, Illinois, USA.}\\
\llap{$^{~2}$}{Now at University of Glasgow, UK.}\\
\llap{$^{~3}$}{Also supported by LPNHE, Paris, France.}\\
\llap{$^{~4}$}{Now at Imperial College, University of London, UK.}\\
\llap{$^{~5}$}{ITEP, Moscow, Russian Federation.}\\
\llap{$^{~6}$}{Now at SpinX Technologies, Geneva, Switzerland.}\\
\llap{$^{~7}$}{Now at TRIUMF, Vancouver, Canada}\\
\llap{$^{8}$}{On leave of absence from Ecoanalitica, Moscow State University,
Moscow, Russia.}\\
\llap{$^{9}$}{Now at \c{C}ukurova University, Adana, Turkey.}\\
\llap{$^{10}$}{Now at III Phys. Inst. B, RWTH Aachen, Aachen, Germany.}\\
\llap{$^{11}$}On leave of absence from INFN, Sezione di Ferrara, Italy.\\
\llap{$^{12}$}{Now at CERN, Geneva, Switzerland.}\\
\llap{$^{13}$}{Now at Univerity of Rome Tor Vergata, Italy.}\\
\llap{$^{14}$}{Now at Lawrence Berkeley National Laboratory, Berkeley, California, USA.}\\
\llap{$^{15}$}{MiniBooNE Collaboration.}\\
\llap{$^{16}$}{Now at Section de Physique, Universit\'{e} de Gen\`{e}ve, Switzerland, Switzerland.}\\
\llap{$^{17}$}{Now at Royal Holloway, University of London, UK.}\\
}
\abstract{
In the HARP experiment the large-angle spectrometer is using a cylindrical TPC as main
 tracking and particle identification detector.
The momentum scale of reconstructed tracks in the TPC is the most important
systematic error for the majority of kinematic bins used for 
the HARP measurements of the double-differential production cross-section 
of charged pions in proton interactions on nuclear targets at large angle.
The HARP TPC operated with a number of hardware shortfalls and
operational mistakes.
Thus it was important to control and characterize its momentum calibration.
While it was not possible to enter a direct particle beam into the
sensitive volume of the TPC to calibrate the detector, a set of physical
processes and detector properties were exploited to achieve
a precise calibration of the apparatus. 
In the following we recall the main issues concerning the momentum measurement
 in the HARP TPC, 
and describe the cross-checks made to validate the momentum scale.  
As a conclusion, 
this analysis demonstrates that the measurement of momentum 
 is correct within the published precision of 3\%.
}
\begin{document}

\section{Introduction}
\label{sec:Introduction}

The HARP experiment~\cite{ref:HARP-prop,Detector-paper} at the CERN PS
was designed to make measurements of hadron yields from a large range
of nuclear targets and for incident particle momenta from 1.5~\GeVc to 15~\GeVc.
The main aims are to measure pion yields for a quantitative
design of the proton driver of a future neutrino factory, 
to provide hadron production cross-sections for precision calculations
of the atmospheric neutrino 
flux~\cite{ref:carbonfw} and to measure particle yields as input for the flux
calculation of accelerator neutrino experiments,
such as K2K \cite{Al-paper}, MiniBooNE and SciBooNE \cite{Be-paper}.

 The HARP experiment makes use of a large-acceptance spectrometer consisting of a
 forward and large-angle detection system.
 A detailed description of the experimental apparatus can be found in reference~\cite{Detector-paper}.
 The forward spectrometer --- 
 based on large area drift chambers~\cite{ref:NOMAD_NIM_DC} and a dipole magnet
 complemented by a set of detectors for particle identification (PID): 
 a time-of-flight wall~\cite{TOFW-paper} (TOFW), a large Cherenkov detector (CHE) 
 and an electromagnetic calorimeter  ---
 covers polar angles up to 250~mrad which
 is well matched to the angular range of interest for the
 measurement of hadron production to calculate the properties of
 conventional neutrino beams.

 The large-angle spectrometer --- based on a Time Projection Chamber (TPC)
 and Resistive Plate Chambers (RPCs), 
 located inside a solenoidal magnet ---
 has a large acceptance in the momentum
 and angular range for the pions relevant to the production of the
 muons in a neutrino factory.
 It covers the large majority ($\sim 70\%$) of the pions accepted in the focusing
 system of a typical design.

\subsection{The HARP TPC}

The HARP TPC was designed and built in a record time of
about 1.5 years. Its main design features are an almost full solid
angle acceptance and high-event rate capabilities. 
It was operated in the years 2001 and 2002 at the CERN PS. Additional 
specialized calibration runs were performed in 2003. 

\begin{figure}
  \includegraphics[width=0.67\columnwidth]{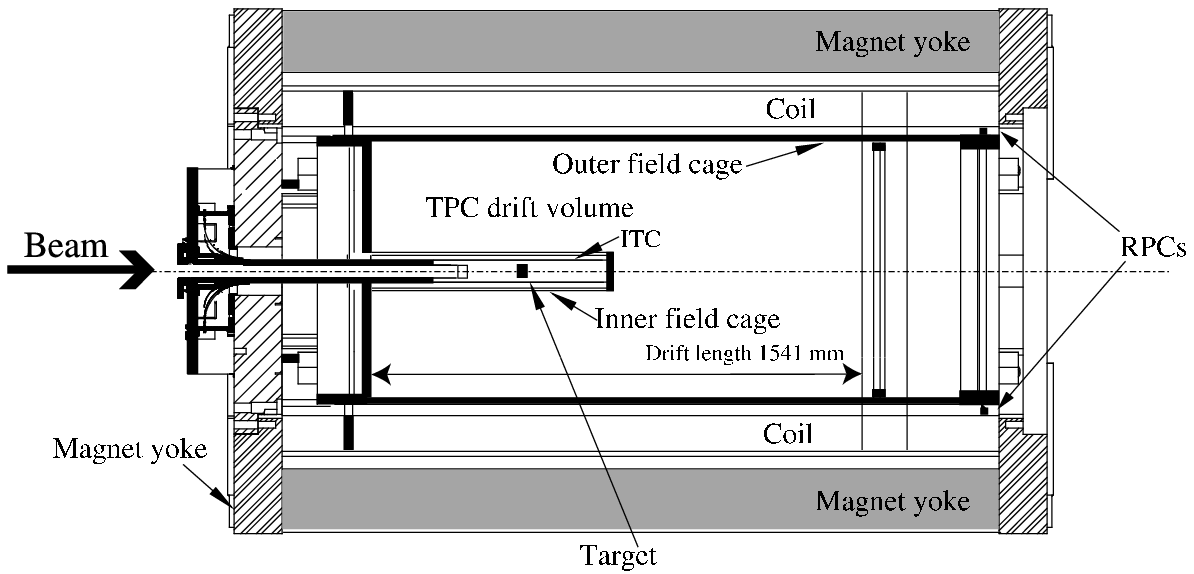}
  \includegraphics[width=0.31\columnwidth]{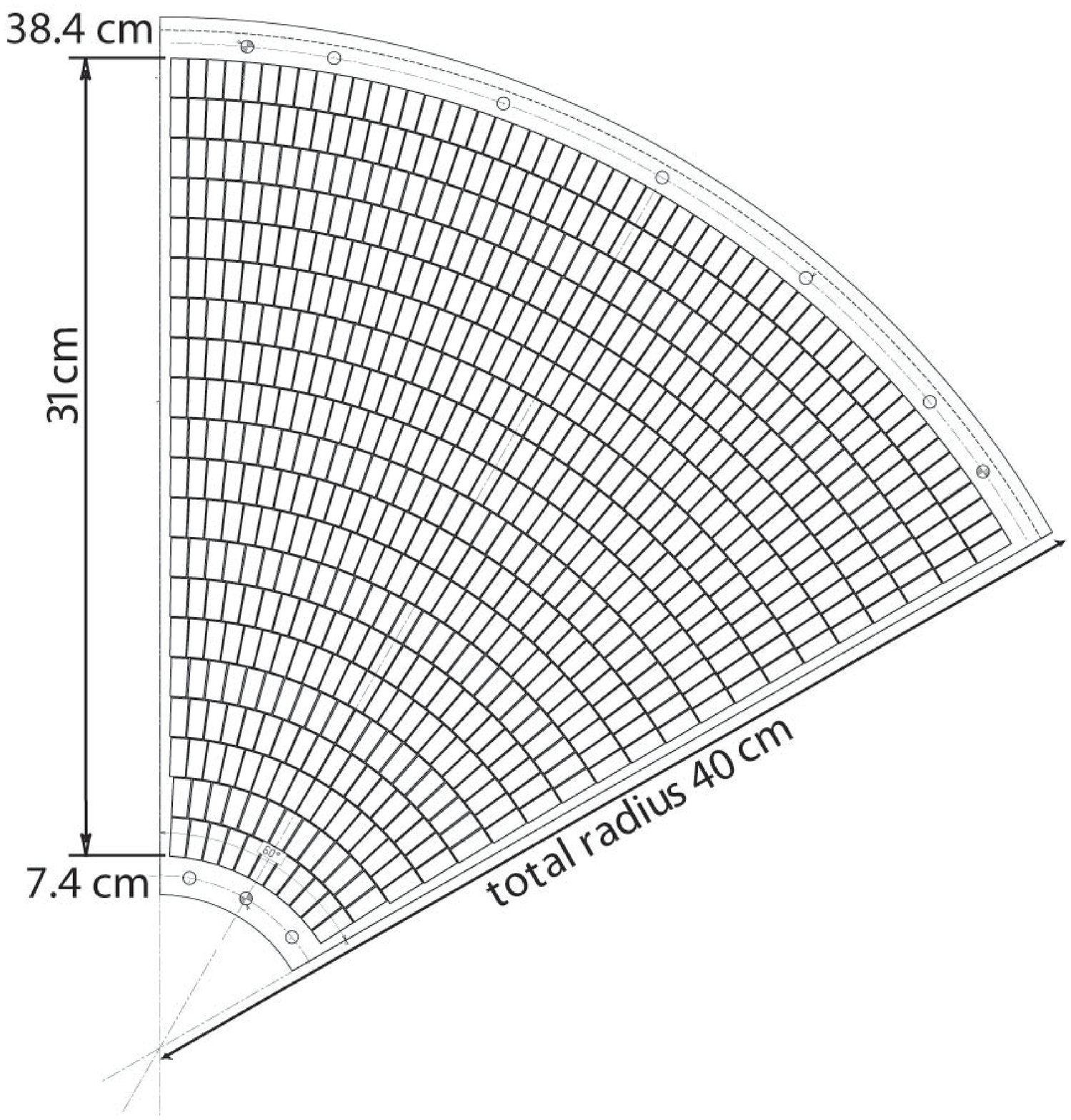}
  \caption{Left panel: schematic layout of the TPC.
   The beam enters from the left. Starting from the outside, first the return
   yoke of the magnet is seen, closed with an end-cap at the upstream end,
   and open at the downstream end.
   The field cage is positioned in the middle of the
   magnetic volume. The inner field cage is visible as a short cylinder
   entering from the left. The ITC trigger counter and target
   holder are located inside of the inner field cage.
    Right panel: mechanical drawing of a sector of the TPC pad plane,
   the layout of the pads is indicated.
  }
  \label{fig:TPC}
\end{figure}

The TPC consists of a cylindrical volume 1.5 m long and 0.8 m diameter
 filled with a 91\% Ar, 9\% CH$_4$ gas mixture  positioned in a solenoidal
 magnet with a field of 0.7 T.
A 12 kV electric field drives the ionization charges at a velocity
of $5 \ \cm/\micros$ to the read-out plane, where the induction signals
 are collected by 3972 pads arranged in 20 concentric rows.
 The pad signals are digitized in 100 ns time bins,
corresponding to about 5 mm bins in the longitudinal direction.
A sketch of the HARP TPC and of its pad plane is shown in Fig.~\ref{fig:TPC}.
More technical details can be found in reference~\cite{Detector-paper}.
The TPC is the key detector for the analysis of 
tracks emerging from the target at large angles with respect to the
incoming beam direction.

 The HARP TPC suffered from a number of shortcomings that were discovered
 during and after the data taking~\cite{Detector-paper}:  
\begin{enumerate}\addtolength{\itemsep}{-0.2\baselineskip}
\item
 A rather large number of deficient electronic channels ($\sim 15\%$)
due to poor soldering of a fraction of cables to the back of the
pad plane.
\item Static distortions caused by the inhomogeneity of the magnetic
  field, the accidental HV mismatch (about 2\%) between the inner and outer field
  cage and  edge effects near the inner and outer field cage.
\item Dynamic distortions caused by build up of ion-charge density in 
  the drift volume during the 400~ms long beam spill, caused by 
  a partial 'transparency' of the cathode wire grid.
  Given the beam intensity and the data acquisition rate with
  the 5\% interaction length targets, it follows that HARP operated under
  conditions of a high dead time (higher than 90\%). 
  The number of events collected in each spill was on average about 300.
\item Cross-talk between pads caused by capacitive coupling between
  signal lines in the multilayer printed boards. 
\end{enumerate}
A description of the measures taken to
correct for the effects of items 1, 2, and 4 is given in~\cite{Detector-paper,
Ta-paper}.\\
The treatment of the dynamic distortions and some detail of the track fitting procedure are described in Appendix~\ref{Appendix}.
We recall here that in large angle cross section results published so far
\cite{Ta-paper,ref:HARP:carbon}, only the first part of the spill (about
30\% of the total events), where the dynamic distortions are negligible
were used (as discussed in Appendix~\ref{Appendix} the distortions can be monitored by a physical parameter named $d'_\circ$).
This provides very little penalty in measuring cross sections because already with this statistics systematic errors dominate in most of kinematic bins \cite{Ta-paper,ref:HARP:carbon}.

Under these experimental conditions,
in the absence of an appropriate calibration system and without the possibility
of exposing the TPC to test-beams,
 a wide range of experimental cross-checks
has been employed to assess the momentum scale in the HARP TPC, as
described in the following.\\ 

\subsection{Procedure used to determine the absolute calibration of the
 momentum scale }

The momentum measurement in the HARP TPC is a direct
result of the calculation based on the measured track curvature and the
 known magnetic field,
no {\it ad hoc} correction factor has been applied
 to make the measurement agree with the benchmarks.
Thus, the determination of the scale should be considered as a cross-check
 rather than a calibration.

A bias on the momentum scale as measured by a TPC is typically
related to a sagitta error:
\begin{equation}
   \delta(\pt)/\pt=s\cdot 8 \cdot q \cdot \pt /(0.3 \cdot B \cdot L^2)
   \label{eq:sagitta}
\end{equation}
Where the sagitta $s$  and  the track length $L$ are in meters (0.5 m is
the typical track length in HARP),
 the magnetic field $B$ in Tesla (0.7 T in HARP),
the track momentum \pt is in \GeVc, $q$ is the sign of the charge
of the particle.

Unfortunately, it was not possible to send a direct beam of particles
into the sensitive volume of the TPC.
In the absence of such a beam, well defined procedures were used to determine
the absolute calibration of the absolute momentum calibration of the TPC.
\begin{itemize}
\item
The momentum scale in the TPC was characterized by using proton--proton
elastic scattering data as benchmark, see Section~\ref{sec:elastics},
 in two different ways:
 \begin{enumerate}
  \item
  By using  the incident proton momentum and direction (measured
  by the beam MWPCs) and the momentum and direction of the proton
  scattered at large angle, measured by the TPC,
  the missing mass squared $M_x^2$ is determined for every event 
  (see Section~\ref{sec:missing-mass}). 
  A bias in the momentum scale would reflect in a bias in the $M^2_x$ calculation.
  \item
    The  angle of the forward scattered particle is used
    (measured by the forward spectrometer) together with the momentum
       and direction of the incoming proton  to
    predict from the kinematics of the elastic scattering
    the recoil proton momentum and direction.
    This prediction is then compared with the measured momentum
    of the recoil proton (see Section~\ref{sec:direct-comparison}).
    Special care has been devoted in this test
    to avoid any bias due to the different
    energy losses of protons (measured in elastic scattering events) against
    pions (cross section measurements), as
    described\footnote{In principle it is enough to measure the angle of the scattered
    proton to predict its momentum. We did not follow this method because
    a) we use the angle to select a clean sample of elastic scattering
    events and b) the angle of protons is affected by multiple scattering
    in the material around the target} in Section~\ref{sec:unconstrained}.
 \end{enumerate}
  \item
     As an additional cross-check, one can also look at the \dedx distribution,
     see Section~\ref{sec:dedx}.
     A satisfactory description of the $p$--\dedx distribution is 
     obtained after the TPC calibration.  
     Although  less precise than the elastic scattering
     kinematics this method can be used to exclude large biases.
\item
     A sagitta error would have opposite sign for positively and
     negatively charged particles and would grow linearly with
     \pt. 
     It would thus be detectable, regardless of the absolute scale,
     by a dependence of the measured total momentum on the track angle
     for samples of tracks with different angles for which one can
     ensure that they have the same total momentum.
     These samples, as discussed in Section~\ref{sec:sagitta}, 
     can be defined using protons in fixed regions of
     relatively high \dedx  (\dedx depends only on the total momentum).
\item
     The $p$--$\beta$ relation using the time-of-flight measurement with
     the RPCs can also be used as a
     relatively weak cross-check, see Section~\ref{sec:rpc}.  
     The precision of this method is limited by the understanding of
     the detector physics of the RPCs in combination with the very short
     flight-path. 
\end{itemize}

\section{Elastic scattering data}
\label{sec:elastics}
\subsection{Measure of the missing mass squared\label{sec:missing-mass}}
   This analysis has been already published in \cite{Detector-paper} and it is only briefly summarized here.
   Events from the 3~\GeVc momentum runs are selected by requiring standard beam selection criteria for protons and
   only 1 or 2 prong events in the TPC. 
 The 2-prong events are determined by very loose kinematical cuts:
$\vert (\phi_1-\phi_2 ) - \pi \vert <0.3$ rad  and
$(\theta_1 + \theta_2) < 1.75$ rad,
 where $\phi_1,\,\phi_2,\,\theta_1,\, \theta_2$  are,  respectively,
 the azimuthal and polar angles of the two tracks.
Further selection criteria are applied to the large-angle track, that is used for the final analysis: the particle is positively charged and well measured over a minimum of 10 points; the reconstructed momentum is in the range 
$320 \ \MeVc \ \le \ p \ < \ 620 \ \MeVc$.  
The tracks must come from the target\footnote{Longitudinal position of
   the point of minimum distance between the beam axis and the track
   extrapolation in the direction of the interaction vertex must be in
   the range of $-50 \ \mm \ \le \ z < \ 70 \ \mm$, where $z$ is the
   coordinate along the beam direction}
and must be recognized as a proton with a \dedx selection.

The missing mass is then computed as:
\begin{equation}
  M^2_x = (p_{\rm beam} + p_{\rm target} - p_{\rm TPC})^2
\end{equation}
where
   $p_{\rm beam}$,  $p_{\rm target}$ , $p_{\rm TPC}$ are the 4-momenta of the incoming beam particle, target particle and the particle scattered at large angle and measured in the TPC, respectively.

\begin{figure}
  \centering
  \includegraphics[width=0.49\columnwidth]{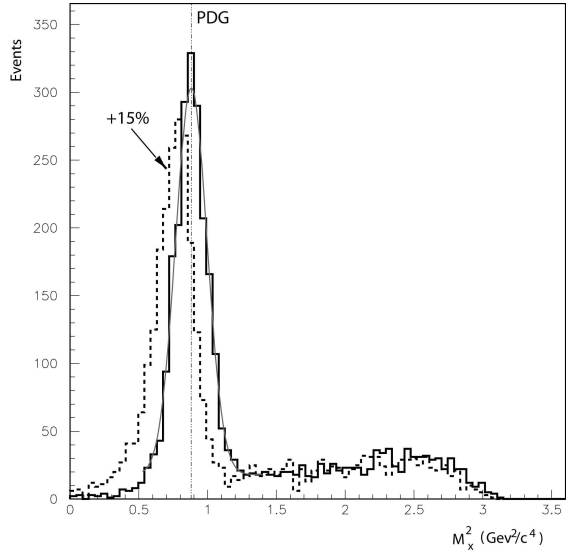}
  \caption{ Missing mass in 3 \GeVc pp scattering.
    The result (solid line) is centered very
   close to the PDG value of the squared proton mass.
   An artificial shift of 15\% of the momentum measured was applied to
   obtain the dashed histogram.  Such a shift is clearly excluded by the
   data (see the text).  }
  \label{fig:mix}
\end{figure}

   The result of this analysis is shown in Fig.~\ref{fig:mix}.  
   A fit to the distribution of Fig.~\ref{fig:mix}
   provides $\langle M_x^2 \rangle = 0.8819  \pm 0.0032  \; \mathrm{GeV^2/}c^4$
   ($\chi^2/{\rm ndof}=20.5/17$ in the $0.55-1.4  \; \mathrm{GeV^2/}c^4$ range
   for a fit using a Gaussian plus a linear background as description) 
   in agreement with the PDG value of $0.88035   \; \mathrm{GeV^2}/c^4$.

   To study the effect of a momentum scale bias over the reconstructed missing mass, we have reconstructed the same distribution by displacing the momentum of the reconstructed track by 15\%.
   As shown in Fig.~\ref{fig:mix} such a bias would produce
   a displacement of about $0.085  \; \mathrm{GeV^2/}c^4$ on $M_x^2$.

   Systematic errors to this measurement come from uncertainties on the primary beam
   particle momentum, correction for proton energy losses 
   in the material of the cryogenic target and inner field cage.
   As a result,  the momentum
   scale is estimated to be correct to better than 3.5\%
   (at one standard deviation).
%
\subsection{Comparison of the measured proton momentum with the elastic scattering predictions\label{sec:direct-comparison}}
%
Elastic scattering interactions of protons and pions off hydrogen
provide events where the kinematics is fully determined by any of the
kinematic quantities and in particular by the
direction of the forward scattered beam particle.
These kinematic properties of the elastic scattering reaction
were exploited to provide a known `beam' of protons pointing into the
TPC sensitive volume. 
Data were taken with liquid hydrogen targets at beam momenta of
3~\GeVc, 5~\GeVc and 8~\GeVc. 

\subsubsection{Data selection}
A good fraction of forward scattered protons or pions  in the elastic
scattering reaction enter into the acceptance of the forward
spectrometer (about $50$\% depending on the beam momentum).

Both direction and momentum of the recoil proton are then predicted.

Selecting events with one and only one track in the forward spectrometer and
requiring that the measured momentum and angle of the forward track are
consistent with an elastic reaction already provides an enriched sample
of elastic events. 
To be counted, tracks need not to be inside the acceptance of the dipole
magnet, but need only to be detected in the upstream drift chamber which
covers the full acceptance of particles exiting the aperture of the
solenoid magnet which houses the TPC.
By requiring that only one barrel RPC hit is recorded at the
position predicted for an elastic event 
(the precision of the prediction from the forward spectrometer is
within the RPC pad size)
and within a time window consistent with a proton time-of-flight, 
a sample of recoil protons with known momentum vector is obtained with a purity of about 99\%. 

The requirement of one RPC hit is relaxed for events where the
recoil proton momentum is predicted to be low enough that it can be
absorbed in the material in front of the RPCs.
In such cases also events without any RPC hit are accepted.
The additional requirement that the recoil angle is consistent with
elastic scattering is then used to ensure a pure sample.
At beam momenta in the range 3~\GeVc--8~\GeVc the kinematics are such
that these protons point into the TPC with angles of $\approx
70^{\circ}$ with respect to the beam direction.

The correlation of the forward scattering angle and
recoil proton momentum introduces an unavoidable threshold in recoil
proton momentum ($\approx 350 \ \MeVc$) which translates into a minimum
angle for the scattered particle. 
The threshold is relatively high due to the need to detect the proton
also in the barrel RPC system outside the outer field cage of the TPC.
As mentioned above, this requirement can be removed only in cases where
a somewhat larger background can be tolerated.

Due to the geometry of the rectangular aperture of the dipole magnet
of the forward spectrometer
only two small horizontal sectors of the TPC can be populated with
recoil protons above threshold momentum in the 3~\GeVc beam.  
In the 5~\GeVc beam the situation is much better and all azimuthal
angles can be populated, although not yet homogeneously.
In the 8~\GeVc beam the population is homogeneous in $\phi$, but the
error propagation of the measurement of the forward scattering angle
into the prediction of momentum and angle of the recoil proton becomes
less favorable.

The numbers of selected elastic events amount to about 15,000 for the
8~\GeVc data sample, and 5,000 for each of the 5~\GeVc and 3~\GeVc data
samples. 
The exposures with higher momentum beams have not been used for this
study. 

%
\subsubsection{Protons versus Pions}
\begin{figure}
  \centering
  \includegraphics[width=0.95\columnwidth]{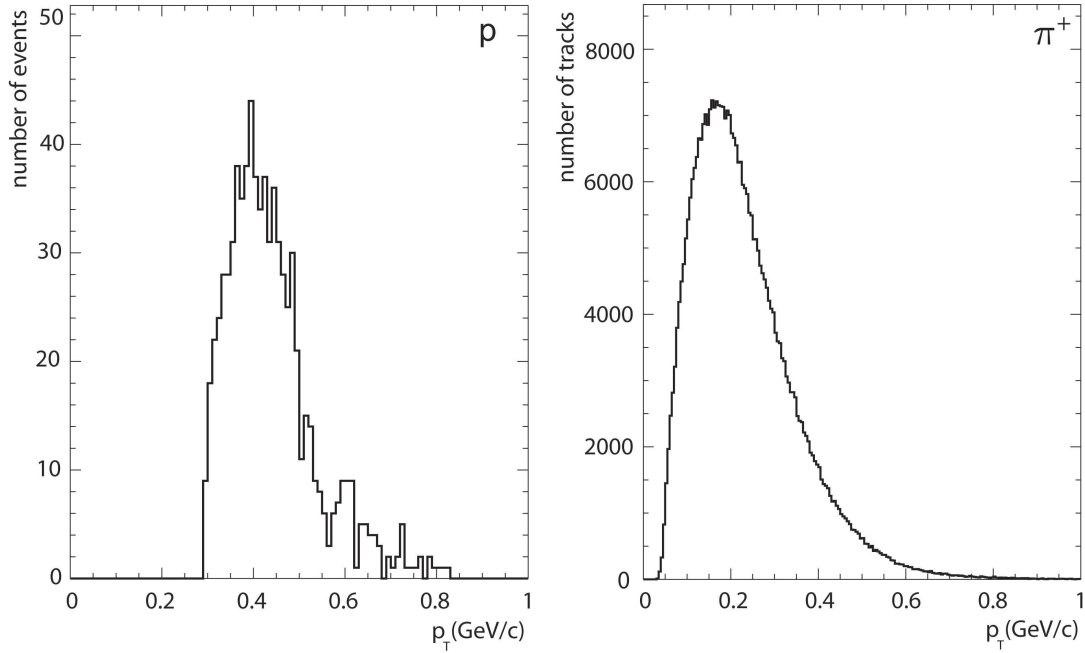}
  \caption{
Left panel: \pt of the recoil protons in used in the proton and pion
 elastic scattering  data (5 \GeVc runs) using the forward spectrometer to determine the
 kinematics. 
  Right panel: typical distribution of the \pt of pion tracks used 
  in the cross section measurement for 8.9~\GeVc p--Be interactions in the
 angular range of the analysis before $p$ and \pt cuts.} 
  \label{fig:confr-pt}
\end{figure}

%
With elastic scattering we can check the reconstructed momentum of protons, 
while in cross section measurements we are interested in the momentum of
pions.
If the momentum scale is influenced by a bias, however, protons are a
robust check provided that their momentum in elastic scattering events
is similar to the momentum of pions in cross section measurements and
that their higher energy losses  do not influence the measurement.
The comparison of the \pt of protons from elastic events and the
\pt of pions in a typical setting, Fig.~\ref{fig:confr-pt}, shows that with elastic scattering
most of the range of interest is covered.

The possibility that the energy loss of low momentum protons can 
alter the momentum reconstruction is discussed in the
following section.
%
\subsubsection{The ``unconstrained fit''\label{sec:unconstrained}}
%
Since the energy loss in the material of the cryogenic target, trigger
counter, and inner field cage is large for protons in the energy range
covered by elastic scattering, there is a significant change of
curvature of their trajectory in that region of the detector. 
This effect introduces a bias in the measurement of the momentum
if one uses the vertex constraint for these low-momentum protons.
Therefore, the behaviour of the momentum measurement for protons was
studied without making use of the vertex constraint.   
If one would use a vertex constraint in the fit for these protons one
would either have to modify the algorithm to take into account the
change of curvature induced by the large energy loss in the inner field
cage or one would have to correct {\em a posteriori} for the bias.
The former option, the use of a modified algorithm, would not validate
the standard code used for the minimum ionizing pions.
The latter option is used in the analysis described in
Section~\ref{sec:missing-mass}. 
Inside the TPC gas volume the energy losses of protons are negligible so
that they can indeed be used to validate the procedures in a way also
applicable to the situation for pions.

Constrained and unconstrained fits are sensitive in the identical way
to any sagitta error, since the vertex  position is not influenced by
distortions in the TPC.

\begin{figure}[tb]
  \begin{center}
    \epsfig{figure=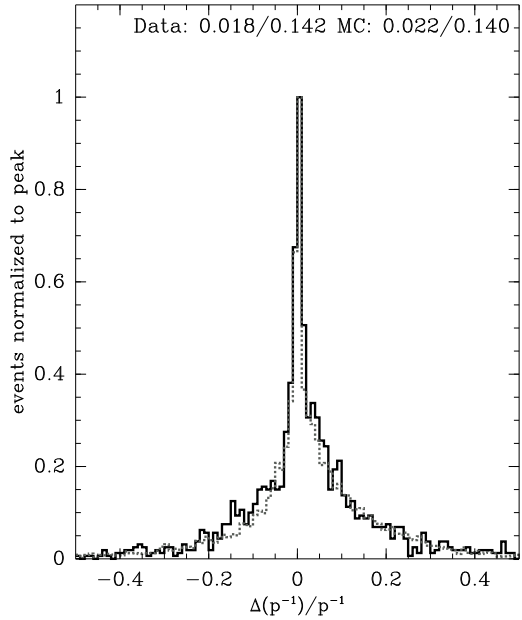,width=0.45\textwidth,angle=0}
  \end{center}
\caption{ 
 Comparison of the unconstrained ($p$1) and constrained ($p$2) momentum
 ($p1/p2-1$) for pions (above 350~\MeVc) using data (from different target materials) 
and the corresponding Monte Carlo.
The data are indicated by the black histogram and the Monte Carlo by the
 dashed histogram.
The position of the peak is at zero well within 1\% and the mean is 2\%
 both for data and MC. 
The
first 50 events in the spill are used.} 
\label{fig:pions-fits}
\end{figure}

For pions and high momentum protons it was checked independently that
the constrained fit is 
unbiased with respect to the unconstrained fit for tracks
reconstructed in the real data and in the simulated data. 
In Fig.~\ref{fig:pions-fits} it is shown that the vertex constraint
does not introduce biases for those particle trajectories and that the
simulation provides an excellent description of the behaviour of the
resolution function.
The comparison of the unconstrained ($p$1) and constrained ($p$2) momentum
 ($p1/p2-1$) for data and Monte Carlo shows 
that the  position of the peak is centered at zero well within 1\% and
that the average is about 2\%  both for data and MC.

%
\subsubsection{Results with the standard data selection}
%
In this comparison, only the first 50 events in the spill were used in
order to avoid the effect of dynamic distortions in the unconstrained
fit (see also Appendix~\ref{Appendix}).
 Given the beam conditions of the run under
study here, this condition guarantees the same data quality as in the
analyses of references~\cite{Ta-paper, ref:HARP:carbon}.

\begin{figure}[tb]
  \begin{center}
    \epsfig{figure=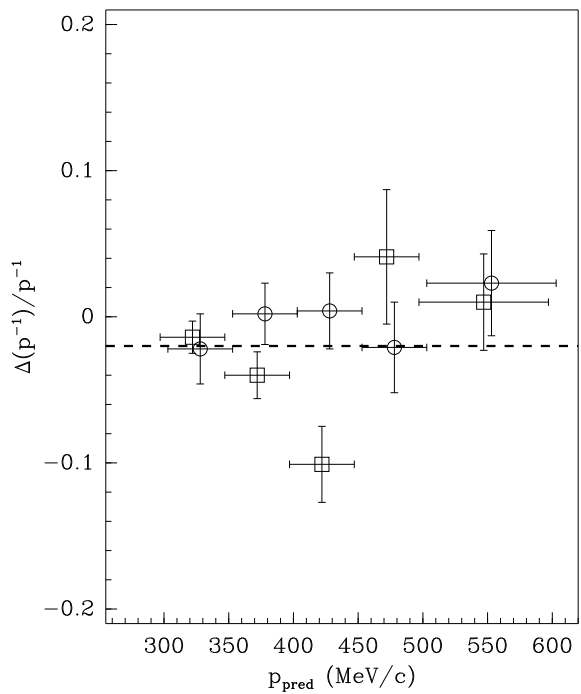,width=0.50\textwidth,angle=0}
  \end{center}
\caption{ 
The momentum bias of the fit without vertex constraint measured with
 elastic scattering data (3~\GeVc: open squares,  5~\GeVc: 
open circles) as a function of the momentum predicted by the forward
scattered track.
In the absence of a clear trend, the average of the points constrains
the bias to be smaller than 3\%.
For these comparisons only the first 50 events in the spill are
used since the unconstrained fit is sensitive to dynamic distortions
beyond this value.  } 
\label{fig:el-p-mom-bias}
\end{figure}

The comparison of predicted momentum and the momentum reconstructed
 without vertex constraint is shown as
a function of predicted momentum in Fig.~\ref{fig:el-p-mom-bias}.
The relative average difference is ($2\pm1$)\%, and shows no clear
 momentum dependence. From this observation one concludes that the
 momentum scale is known to better than 3\% (at one standard deviation).
Systematic uncertainties such as the absolute beam momentum scale,
the precision in the measurement of the kinematic quantities of the
forward scattered track and the need for energy loss corrections limit
this test to a precision of about 2\%.  
Since the sensitivity of the benchmark is similar to the shift observed 
it is not justified to adjust the momentum scale to the benchmark.
%
%
\subsubsection{High statistics benchmarks}
\begin{figure}[tb]
    \includegraphics[width=0.3\textwidth]{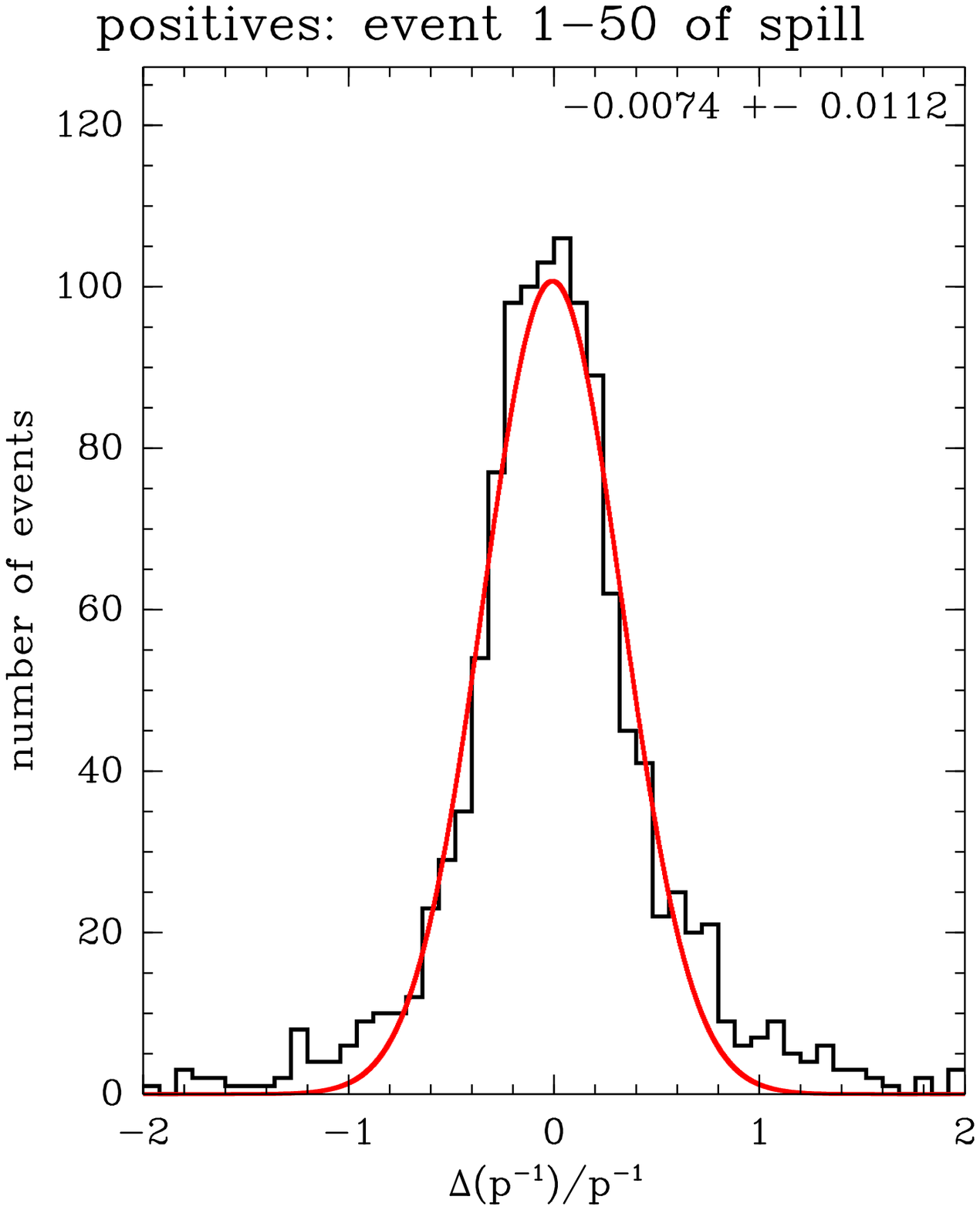}
    \includegraphics[width=0.3\textwidth]{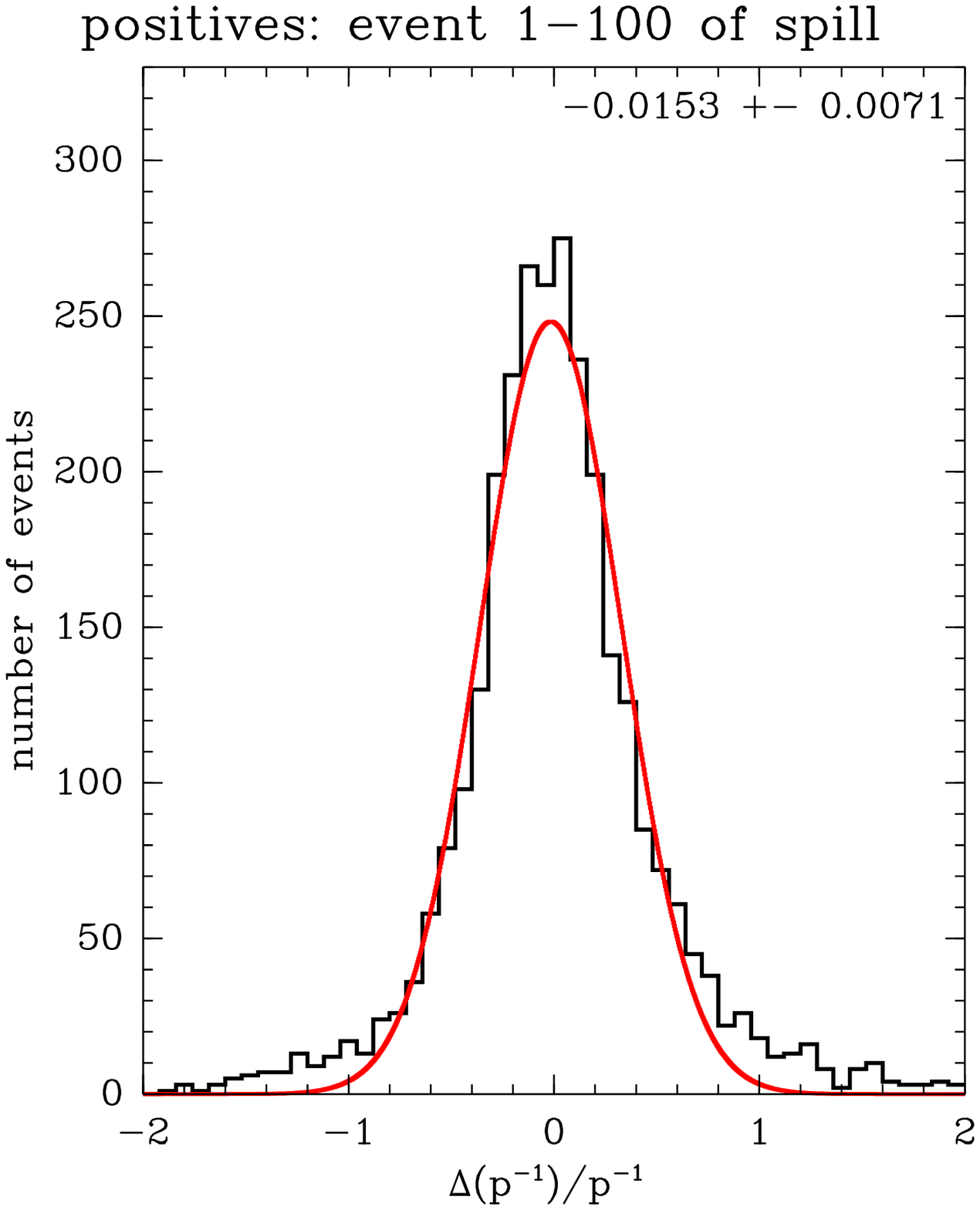}
    \includegraphics[width=0.3\textwidth]{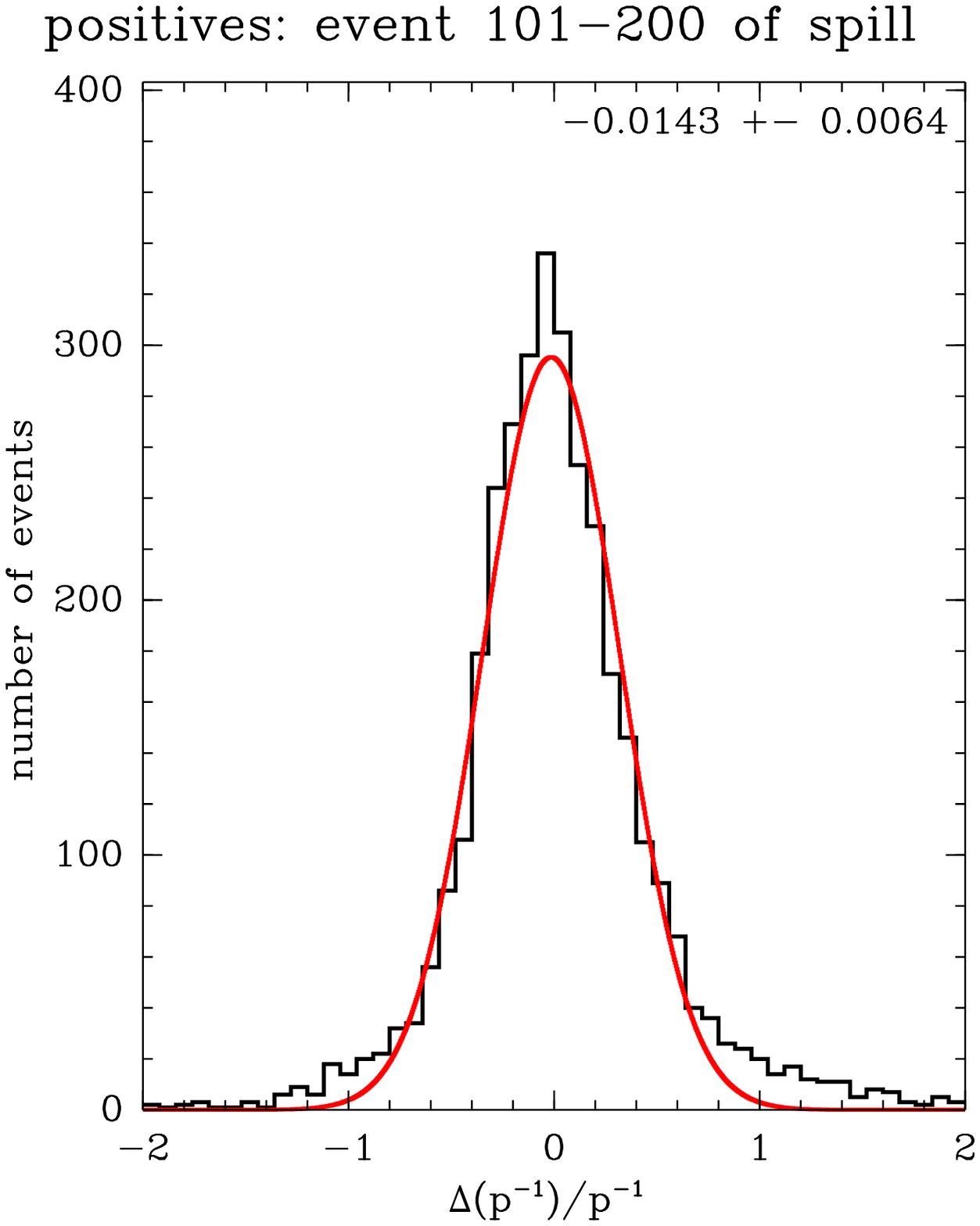}
 \caption{ $\Delta (p^{-1})/p^{-1}$ plot for protons produced in \pip\mbox{p}
  and \mbox{p}\mbox{p} elastic scattering, combining data from the 3, 5 and
  8 \GeVc primary beam momenta.
 Left panel:  first 50 events in the spill, no corrections for dynamic
 distortions.
 Central panel:  first 100 events in the spill, with corrections for dynamic
 distortions.
 Right panel: events 101--200 in the spill, with corrections for dynamic
 distortions.}
 \label{fig:stdvsdyn}
\end{figure}

 To improve the statistics of this check, we make use of the full 
 statistics by applying the correction of the dynamic corrections
 (see Appendix~\ref{Appendix}) and we add elastic scattering
 $\pi^+ - p$ events to the proton elastic scattering sample
 and analyse separately $\pi^- - p$ events
 \footnote{In the following figures the label ``positives''
 indicates the recoil protons in elastic scattering events in the
 positive beam, and ``negatives'' is used to label the protons in the
 negative beam.}. 
 One should note that the effect of a trajectory distortion
 creates the same momentum shift if a systematic shift on the sagitta
 is caused by an $E\times B$ effect, since both the effect and the
 curvature for protons change sign simultaneously. 
Therefore these two settings are expected to provide consistent results.
 The difference between  the predicted and the measured
 $1/p$ (after corrections for the energy loss of the proton prior to entering
  the TPC), is shown in Fig.~\ref{fig:stdvsdyn}.
As mentioned above, this procedure has an intrinsic 2\% systematic error
 coming from the determination 
 of the incoming beam momentum and from the angle measurement with 
 the forward spectrometer.
\begin{figure}[tb]
     \centering
    \includegraphics[width=0.7\textwidth]{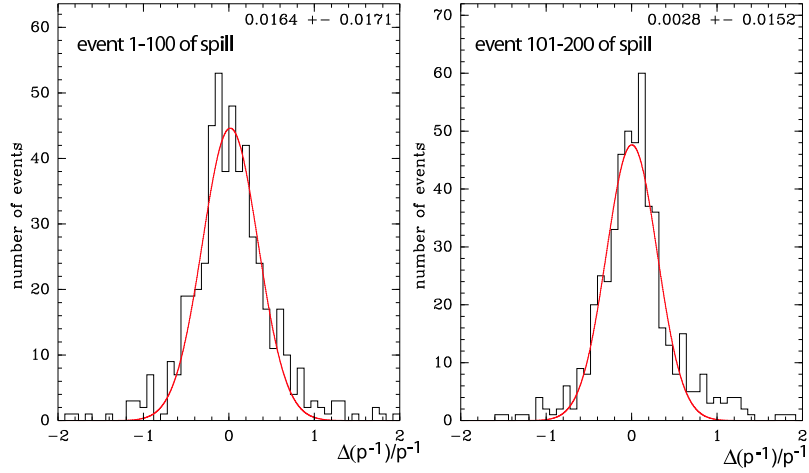}
     \caption{ $\Delta p^{-1}/p^{-1}$ plot for  \pim\mbox{p}
       elastic scattering, combining data from the 3, 5 and
  8 \GeVc primary beam momenta,
       computed with the corrections for dynamic distortions of the TPC.
       Left panel: first 100 events
       in the spill, right panel:  events 101--200 in the spill.  }
    \label{fig:negatives-full}
\end{figure}

The following results were obtained: 
\begin{itemize}
 \item
      The elastic scattering sample using the first 50 events
      (without corrections for dynamic distortions) and the elastic scattering
      sample using the events, corrected for dynamic distortions,
      from 1 to 100 and from 101 to 200 are fully compatible
      (see Fig.~\ref{fig:stdvsdyn});
 \item
      With the larger statistics allowed by the use of 200 events per spill
      it is now possible to compare  ``positives''
      (\pip\mbox{p} and \mbox{p}\mbox{p}) (Fig.~\ref{fig:stdvsdyn})
      and ``negatives''  (\pim\mbox{p}) (Fig.~\ref{fig:negatives-full}).
\end{itemize}
The distribution for ``positives'' has an average $\Delta (p^{-1})/p^{-1}$ equal to $-0.0148 \pm 0.0047$
while the distribution for ``negatives'' has
 $\langle \Delta (p^{-1})/p^{-1} \rangle = 0.0096 \pm 0.0113$.
 The combination of the two polarities gives
$\langle \Delta (p^{-1})/p^{-1} \rangle = -0.011 \pm 0.004$. Again, taking into account the systematic
errors, we conclude that no bias on momentum is observed with a precision of
3\%.
%
\subsubsection{Stability of the elastic results with other settings}
%
To check that the results obtained with the elastic events on the hydrogen target are stable in the other data taking settings, we have selected a narrow
\dedx region corresponding to 7--8 MIP. In this region the pion contamination
is negligible and protons have an average momentum of 340 \MeVc.

\begin{figure}[tb]
  \centering
  \includegraphics[width=0.75\textwidth]{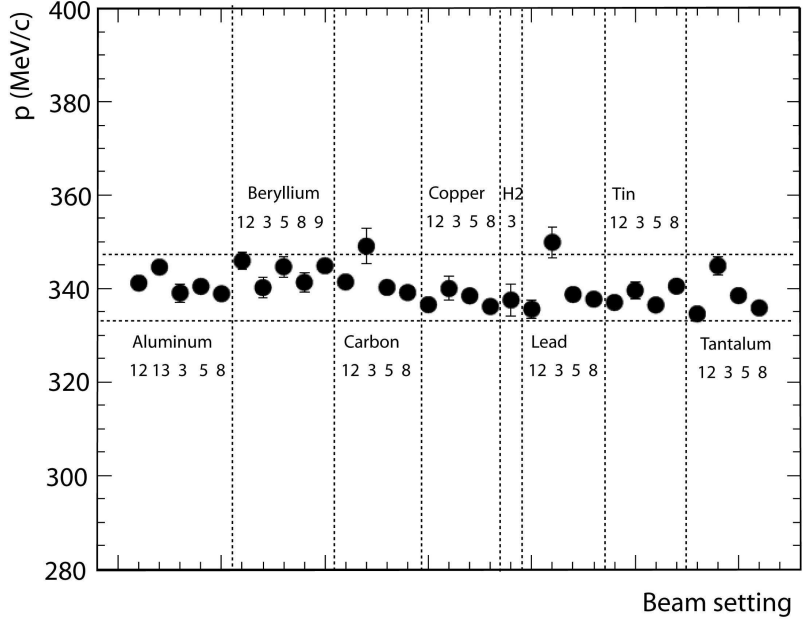}
  \vspace*{-1cm}
  \caption{Average momentum of particles with a \dedx in the TPC corresponding
to  7-8 MIP, as measured in 31 different settings.
The horizontal dashed lines correspond to a variation of $\pm 2\%$ around
the average value of 340 \MeVc. The different settings are labeled with
the material of the target and the momentum, in \GeVc, of the incident beam.
}
  \label{fig:stab-settings}
\end{figure}

The average reconstructed momentum of protons in this band is shown in Fig.~\ref{fig:stab-settings} for 31 different settings. All the settings provide an average momentum
within $\pm 2\%$ around the average value of 340 \MeVc, demonstrating the
stability of the momentum scale measured with elastics during the
overall HARP data taking.
Particles were only accepted when they were nearly perpendicular to the
beam direction, so that the average \pt of this sample is 310~\MeVc.

\section{Track residuals with positive and negative settings }
\label{section:residuals}

A way to monitor the presence of residual distortions
 (when the dynamic distortion correction is not applied)
is to look at the  $\Delta(R \phi)$  difference between the coordinate
 of the track measured in each pad
 row of the TPC and the trajectory estimated by the circular fit.
 To do this we have selected tracks 
(vertex constrained) hitting  the center of the RPC overlap to be
 able to fix an external point. 
The cuts applied in the standard analysis have been used.
The same residual distributions can be obtained  separately for positive 
and negative magnetic field direction.  
In this case an $E\times B$ effect changes sign for the two polarities.
For this test we used a carbon 5\% nuclear interaction length ($\lambda_I$)
 target with beam momenta of $\pm 5$~\GeVc respectively.

\begin{figure}[tb]
  \centering
  \includegraphics[width=0.49\textwidth]{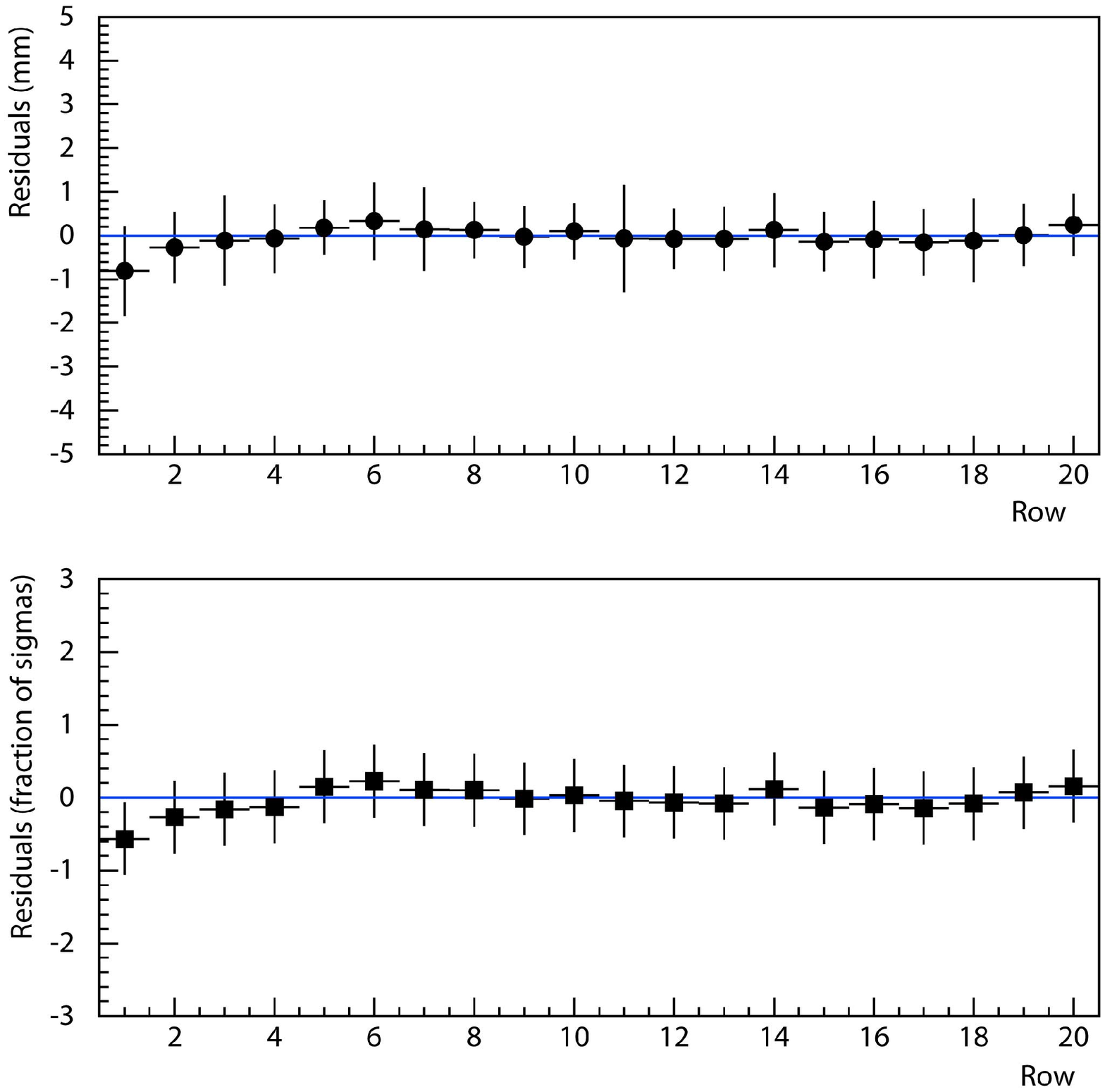}
  \includegraphics[width=0.49\textwidth]{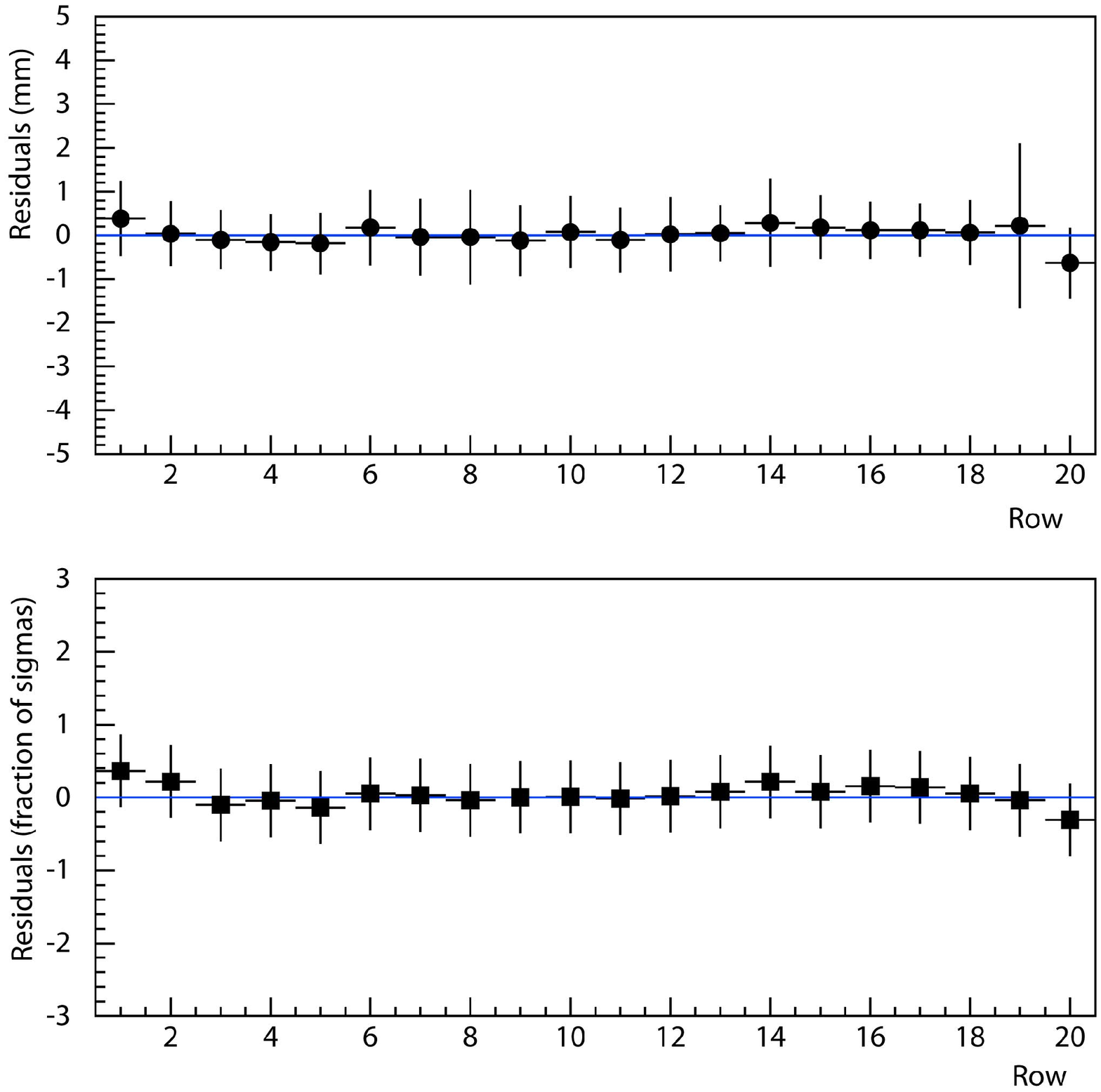}
  \caption{ Left panel: the mean residual $\Delta(R \phi)$ for each pad
   row of the TPC measured using a $B$ field positive polarity setting
 (+5~\GeVc Carbon target data).
  Top: $\Delta(R \phi)$ in mm. Bottom:  $\Delta(R \phi)$ in fraction of RMS .\ 
  Right panel: Same using a  $B$ field negative polarity setting
 ($-5$~\GeVc Carbon target data).\
  Changing the $B$ field polarity, the swap in sign of the mean residuals in the innermost and 
  outermost pad ring is clearly visible }
  \label{fig:res+-}
\end{figure}
The analysis of the distributions of the residuals shows that the biases are
 small (in the range of $\pm 200$ microns).
As expected row number 1 (the innermost) and row number 20 (the
 outermost) display edge effects 
($-$800 $\mu$m and +300 $\mu$m respectively) which are not fully addressed 
by the distortion correction for static misalignment between the
 inner and outer field cage voltages.
The fact that the residual is larger in the inner row and of opposite sign to
 that in the outer row is consistent with the hypothesis that the effect
 is due to a  residual electrostatic field, see Fig.~\ref{fig:res+-}
 (left).

A further confirmation was obtained by looking at the residual distribution
 for the tracks of the $-5$~\GeVc sample where the magnetic field polarity was inverted. 
In this last case the behaviour is the same but the sign of the residual of
the innermost and outermost row is now inverted (+380 $\mu$m and $-$635 $\mu$m respectively),
 see Fig.~\ref{fig:res+-} (right).

By excluding rows 1 and 20 from the fit,
 one can place a limit of less than 1\% on the effect of the residual 
distortion effects on the momentum measurement.
\section{Consistency checks of the momentum calibration with \dedx}
\label{sec:dedx}

%
\begin{figure}[tb]
  \centering
  \includegraphics[width=0.59\textwidth]{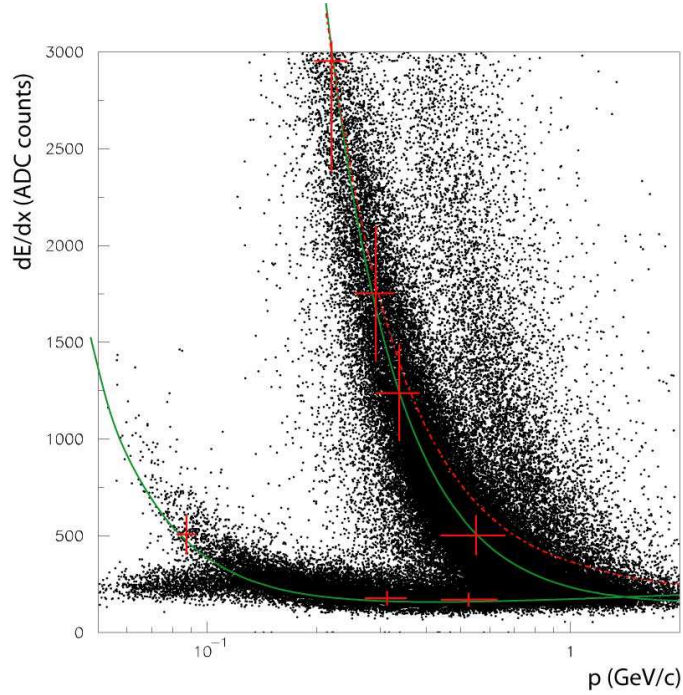}
  \caption{ 
  ${\mbox{d}E/\mbox{d}x}-p$ plot of HARP data, 5\% Ta target
  at 5 \GeVc, fitted with the
  modified Bethe-Bloch
  function (see the text), including the resolution
  bars for every fitted slice in momentum and \dedx.
  The bars are computed from the published momentum resolution and
  \dedx resolution for all points.
  The dashed curve is the $1/\beta^2$ curve.  }
  \label{fig:dedx}
\end{figure}
The \dedx cannot be used in HARP to
estimate the momentum scale with a precision similar to the elastic
scattering method because both the scale and offset
calibration of \dedx are free parameters and the resolution in
\dedx, about 17\%, is insufficient to achieve such a precision.  
Nevertheless, the \dedx--$p$ plot provides a qualitative cross-check 
of the TPC momentum calibration. Indeed we find good agreement
as shown in Fig.~\ref{fig:dedx}.

It has been claimed that the disagreement of the \dedx plots we
published in \cite{Ta-paper} with a $1/\beta^2$ curve is a clear
symptom of a TPC momentum bias, up to 15\% \cite{friendly-comments}.
Since the free parameters of the \dedx curve can only be fixed using the
point at which particles are minimum ionizing, it will be immediately
clear that a $1/\beta^2$ description, which reaches its minimum
asymptotically, cannot be an adequate approximation 
as shown in the comparison of the correct curve and this simple
approximation in Fig.~\ref{fig:dedx}.
Since this was not immediately obvious to the authors of
Ref.~\cite{friendly-comments}, we include here a rather pedantic
discussion of \dedx.
The {\em average} energy loss is described with the standard Bethe-Bloch formula \cite{pdg}:
\begin{equation}
-\frac{dE}{dx}=K z^2 \frac{Z}{A} \frac{1}{\beta^2} \left[
\frac{1}{2}{\rm ln}\frac{2 m_e c^2 \beta^2 \gamma^2 T_{max}}{I^2}
- \beta^2 - \frac{\delta(\beta \gamma)}{2} \right].
\end{equation}
 
For particle identification
a truncated mean is tuned to correctly estimate the Landau
peak position (discarding the 20\% of points with the highest \dedx),
 and not the mean \dedx (for which the standard Bethe-Bloch theory
applies). Hence each point on the \dedx--$p$ scatter-plot represents the
calculation of the most probable \dedx per TPC pad row, integrated over the
tracks' effective path length across each pad row
(therefore it represents the peak value of a convolution of Landau distributions).
Its phenomenology can be described sufficiently accurately by a modified Bethe-Bloch
formula \cite{Bethemod}, as shown in Fig.~\ref{fig:dedx}:
the \dedx  for protons, pions, the  positions of the \dedx of a minimum
ionizing particle (MIP), and intersection points of the bands for
different particle types are all consistent.

\begin{figure}[tb]
  \centering
  \includegraphics[width=0.4\textwidth]{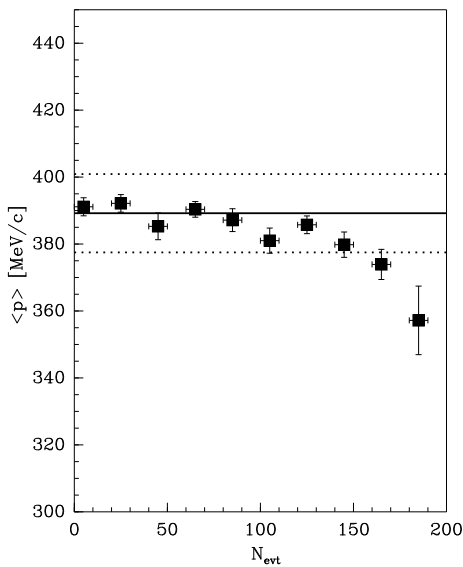}
  \caption{  Average reconstructed momentum as a
    function of event number in spill for protons using a high
   value of \dedx for the selection.
The analysis is performed for the combined data set taken with
3~\GeVc, 5~\GeVc, 8~\GeVc and 12~\GeVc beams on Be, C, Cu, Sn, Ta and Pb
targets. 
   The solid line shows the average for protons for
    the first 100 events in the spill.
   The two dotted lines show the $\pm 3\%$ variation around the
   average. 
}
\label{fig:dedx:pions}
\end{figure}

To avoid the effect of dynamic distortions the above analyses were
     done using only the first 50 events in each spill.
     It was checked that the constrained fit remains stable, well within
     3\%, for about 100 events in the spill as will be described in the
     following section.

\subsection{Stability against dynamic distortions}
One can select samples of tracks with a well defined momentum by
accepting narrow enough \dedx intervals in the region of high values
(the so-called ``$1/\beta^2$'' region).
The \dedx resolution is sufficient to select such a proton sample with
only a 10\% RMS spread in ``true'' momentum.
If the measured average momentum of such samples is compared as a
function of event number in the spill \evtspill strong constraints on the influence
of dynamic distortions on the momentum measurements can be obtained.

In this analysis particles were
selected in narrow bands of \dedx in regions where \dedx depends
strongly on momentum.  
To select a sample with the highest possible momentum, 
the protons were further required to reach the
RPC system (low momentum protons would be absorbed before reaching the
RPCs). 
A further selection 1.0~rad~$<\theta <$~1.5~rad ensures a limited
range of $p_{\mathrm{T}}$.
In addition to a momentum selection also a PID-selection is performed
with the same cuts.
The analysis was performed for the combined data set taken with
3~\GeVc, 5~\GeVc, 8~\GeVc and 12~\GeVc beams on Be, C, Cu, Sn, Ta and Pb
targets. 

The average momentum obtained from a Gaussian fit to the momentum
distribution shows that the average momentum stays constant within a
few percent up to
$\evtspill = 100$ at $p_{\mathrm{T}} \approx 350 \ \MeVc$ 
(see Fig.~ \ref{fig:dedx:pions}).

One observes that the behaviour is not compatible with a linear
dependence as a function of time but the average momentum stays constant
over a long period before a downward trend sets in.
One of the reasons is the fact that the distortion effect does not have
a linear dependence at the beginning of the spill, owing to the fact
that the first ions need to exit the amplification zone before they
distort the field in the drift zone.
This is shown in a little more detail in the Appendix.
This effect ``protects'' the first fifty events in the spill very efficiently.
Another reason for increased stability of the constrained fit under the condition
of distortions is simply that the weight of the vertex constraint compensates
 very well for the distortions, up to the point where, when dynamic corrections
 are not applied, the tracks are so distorted that the reconstruction
 efficiency is affected.

It has been shown with elastic scattering that the absolute track finding
efficiency does not change as a function of event number in the spill.
This result indicates that the distortions are continuous and smooth as
a function of $z$ and $R$.
However, once quality criteria are applied, mainly the requirement that
the tracks emerge from the target, the efficiency is reduced when the
distortions are increasing during the growth of the ion charge.
Since this requirement removes tracks shifting out of the acceptance at
one side, and since the measurements of curvature and of the minimum
distance to the interaction point are correlated, 
the deviation of the average measured momentum from a constant is thus a
single-sided efficiency effect.  

The $p_{\mathrm{T}}$-range covered by this cross-check represents a large range
of the kinematic domain used in the analysis.

%
\subsection{Sagitta errors from momentum-angle correlations\label{sec:sagitta}}
\begin{figure}[tb]
\centering
\includegraphics[width=0.48\textwidth]{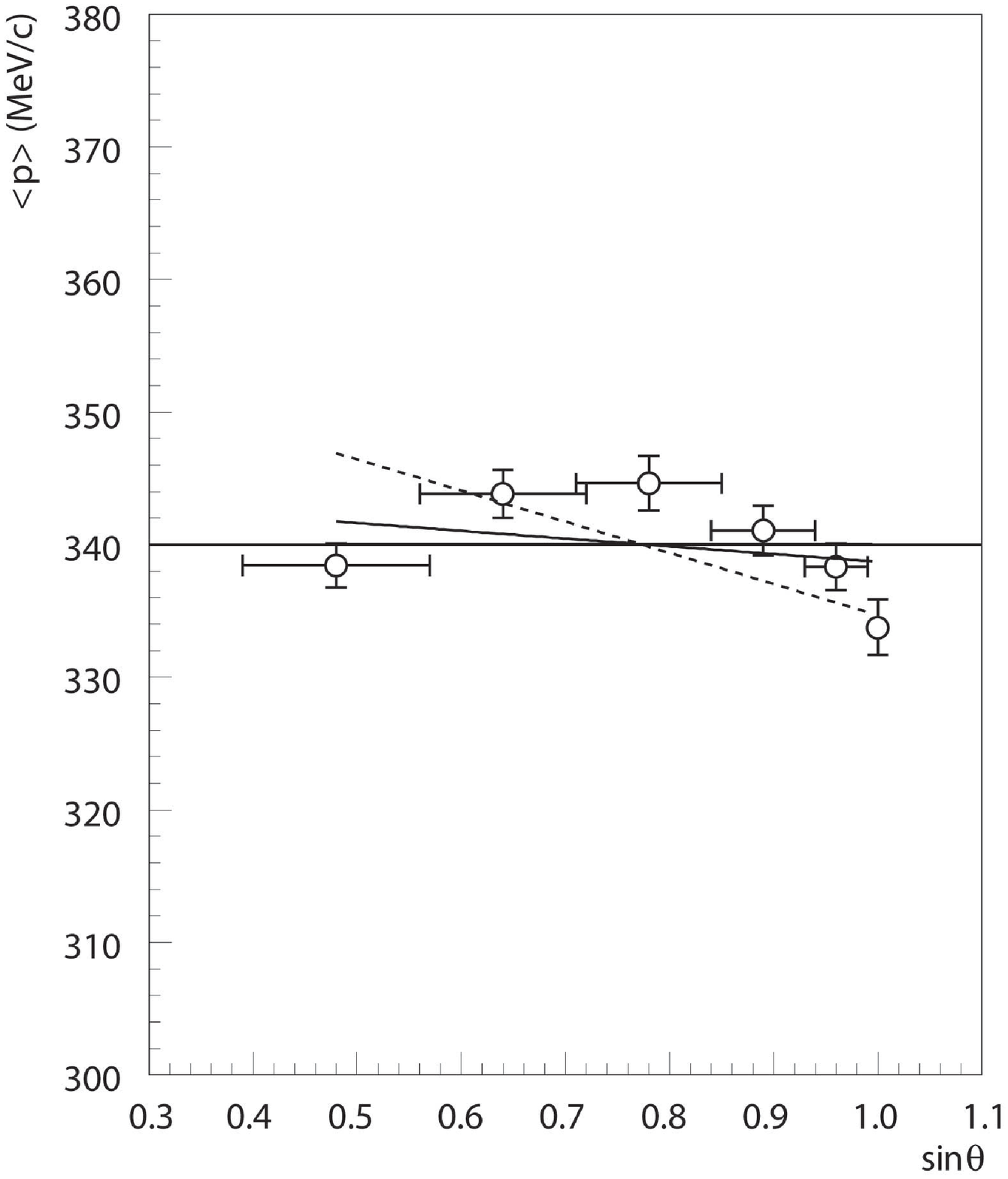}
\includegraphics[width=0.48\textwidth]{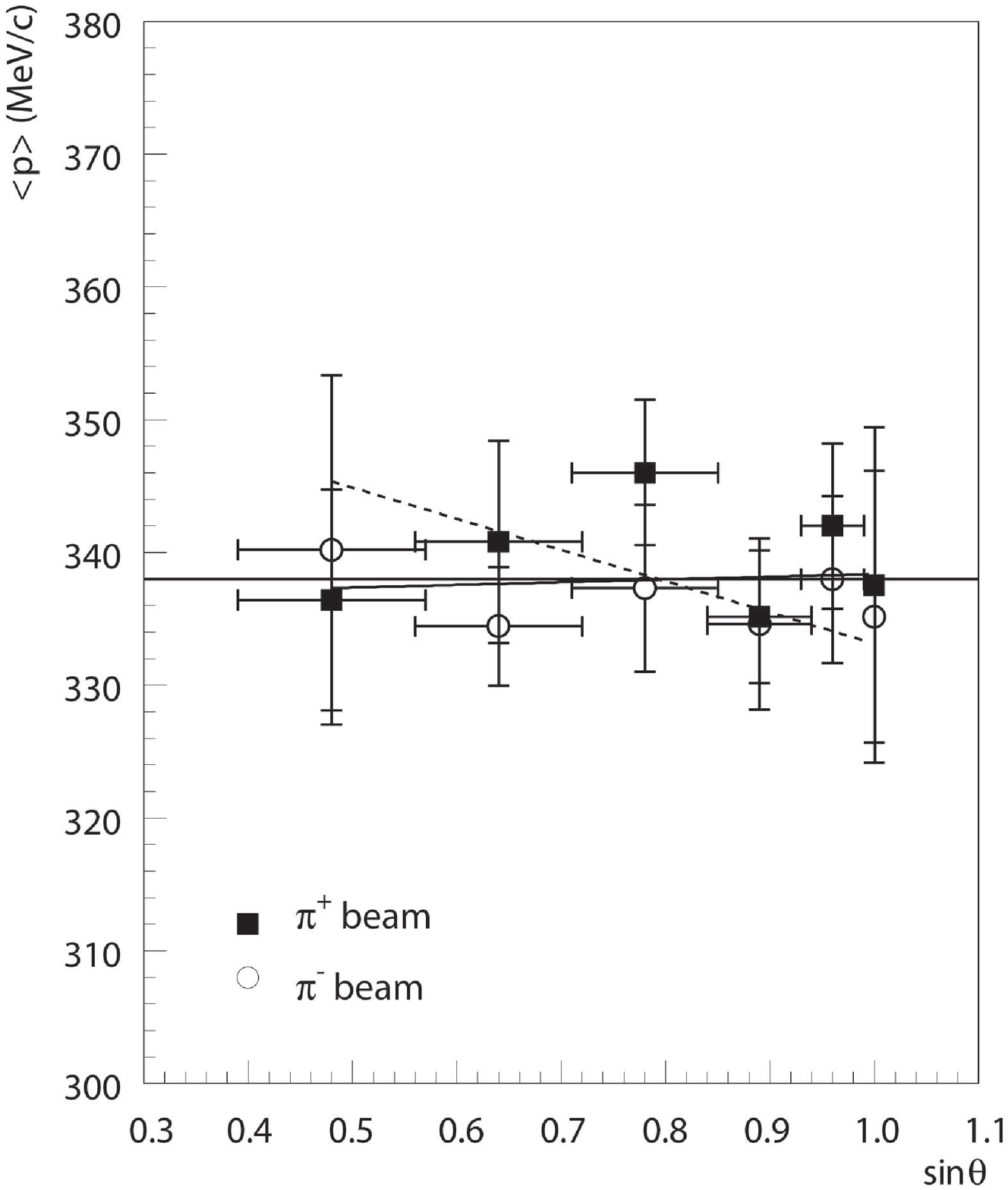}
\caption{  Left panel: average momentum in a fixed slice of \dedx as a function of
 $\sin\theta$.
 Data are collected with Be, C, Cu, Sn, Ta and Pb targets at 3, 5 and 8~\GeVc,
  no correction for dynamic distortions. 
  A fixed shift in sagitta would show up
 as a linear change of average momentum.
 These data have been fitted with a constant term, with a linear function
 (the best fit corresponds to a momentum bias of $\sim 2.5 \%$ at 500~\MeVc)
 and with a linear
 function with a slope corresponding to a 10\% bias (dashed line).
 While the constant term is compatible with the linear function ($\Delta \chi^2=0.8$),
 a 10\% bias has $\Delta \chi^2 \simeq 20$.  Thus, it is excluded
at more than 5 sigma level.
 Right panel: same analysis for
 $\pi^+$ (black squares) and $\pi^-$ (open circles) incident
 beams and with the full spill correction for dynamic distortions.
 These data were taken with opposite magnetic field polarities.
 Data are collected for 8~\GeVc incident beam on Ta target only.
 A fixed shift in sagitta would show up with the same slope
 for positives and negatives.
 In this case, given a smaller statistics, a 10\% bias is excluded
 at about 90\% C.L. ($\Delta \chi^2 \simeq 4.1$).  }
\label{fig:dedx:sin}
\end{figure}

Using a sample of tracks within a fixed interval of \dedx where 
 the average momentum is $\sim 340$~\MeVc, and considering that
$\pt = p_{\mathrm{tot}} \sin\theta$,
it is possible to look for a sagitta bias (acting on \pt)
 through any correlation
between $\langle p \rangle$ and $\sin \theta$.
Unlike previous analyses where the RPC hits were used to set a minimum
 range, here such a requirement was avoided not to introduce an angular
 dependence in the definition of the average energy of the sample.
From the fits to 
Fig.~\ref{fig:dedx:sin} (left) 
we conclude that a null bias is measured with a precision of about 3\%.
%
This analysis has been repeated using
positive and negative pions and the correction for dynamic distortions
with incoming $\pi^+$ in the positive beam
 and $\pi^-$ in the negative beam (Ta target, 8~\GeVc). 
 As shown in Fig.~\ref{fig:dedx:sin} (right),
 for both magnet polarities there is no significant dependence on $\sin
 \theta$.   

Since the curvature of the protons and of distortions (if of the
 $E\times B$ type) are both inverted,
 the slope for protons (if any) is expected to have the same sign for
 positive and negative beams.
The fact that there is no significant dependence on $\sin \theta$
 confirms the reliability of the HARP TPC calibration.

\section{Comparison with time-of-flight measurements}
\label{sec:rpc}

The HARP RPC system \cite{RPC-paper-ieee} is positioned as a barrel around the TPC
 chamber, about 50~cm from the interaction target. It can in principle be used
 to check the momentum calibration comparing the $\beta$--$p$  relation
 of pions and protons, where $\beta$ is measured using the
 time-of-flight to reach the RPC system.

This cross-check is limited in precision due to the short flight distance
of the particles and the rather large corrections needed to convert the
measured threshold crossing time into a measurement of time-of-arrival
of the particle.
For example the range of the correction for the ``time-slewing'' of the threshold
crossing time for different measured integrated charge collected in the
RPCs is 2~ns, similar to the total time-of-flight of pions to reach the
RPCs~\cite{RPC-paper-ieee}.
As an additional complication, the momentum range of the particles for
which a $p$--$\beta$ comparison can be made is in the region where pions
are minimum ionizing and where protons are heavily ionizing (with a
different \dedx by a factor of up to 8).
Thus one first has to ascertain that the response of the RPC system is
well understood before one can use the time-of-flight as a mean to
calibrate the momentum measurement in the TPC.

\begin{figure}
\centering
\includegraphics[width=0.95\columnwidth]{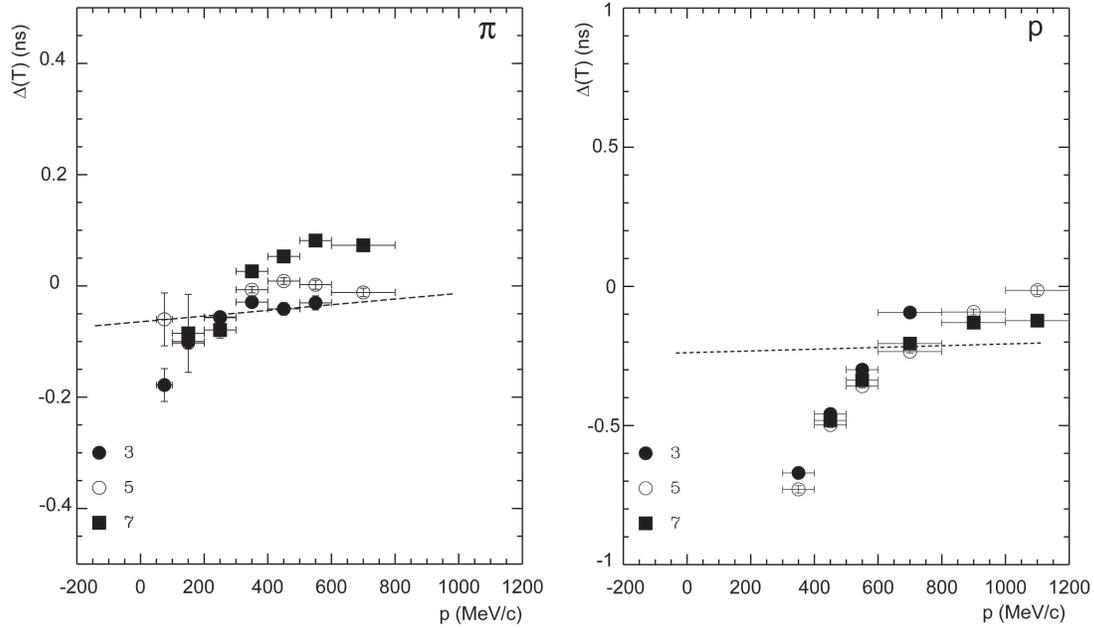}
\caption{ Analysis of $\Delta$ TOF = (measured $-$ predicted) time-of-flight
 for pions (left panel) and protons (right panel).
 The measured time is provided by the
RPC signal time and the predicted time is based on the track momentum measured in the TPC. The
numbers refer to RPC pad ring (equivalent to $Z$ position; with pad 3 in the most backward direction).
Whereas the pion data are centered near zero, the proton data are shifted to
negative times with a positive slope.
The dashed line is the prediction for $\Delta$ TOF for a sagitta bias of 1 mm
and a track length of 0.5 m.  }
\label{fig:RPC}
\end{figure}

Fig.~\ref{fig:RPC} taken from reference~\cite{RPC-paper-ieee} shows the
difference of the time-of-arrival measured with the RPCs, $t_\mathrm{m}$,
and the time-of-arrival predicted using the momentum measured in the TPC,
$t_\mathrm{p}$. 

This plot had been used in \cite{friendly-comments} to claim
a 15\% bias in the HARP TPC momentum scale.

If a momentum bias would be caused by an error in the measurement of the
trajectory sagitta, it would reflect on the $\beta$ of protons and not
on the $\beta$ of pions, which already saturate $\beta$ at
the HARP momenta.
The RPC calibration has been performed using pions, so that one would
expect that these display a vanishing average offset as is the case in
Fig.~\ref{fig:RPC}. 
%
%
However, the behaviour of the measured $\Delta(TOF)$ for protons does not agree
with that predicted by the sagitta model, see Eq~(\ref{eq:sagitta}).
While data, Fig.~\ref{fig:RPC}, exhibits a clear slope,
the sagitta model  predicts a rather flat dependence of $\Delta(TOF)$ on the
measured momentum. This flatness comes from the particular
momentum range of the protons where $\delta(p)/p$
increases linearly with $p$, while $\Delta(TOF)$ decreases with $p$ because
$\beta$ of the protons saturates.

The question whether the RPC time measurement suffers
from systematic effects due to the large difference in primary
ionization caused by pions and protons in the momentum range available
for these calibrations had been addressed with
 a dedicated RPC calibration analysis 
studying proton and pion elastic scattering off the cryogenic hydrogen
target and reported in reference~\cite{phys-RPC}.

As for the measurement of the momentum scale, see Section~\ref{sec:elastics},
such a measurement makes it possible to send a ``controlled beam'' of
slow protons through the TPC and towards the RPC system without the need to
measure the momentum of the recoil proton with the TPC.

An exposure of the HARP detector where a 5~\GeVc beam of
protons and pions is directed onto a 60~\mm long liquid hydrogen
target was used. 

\begin{figure}[t]
\centering
\includegraphics[width=0.50\columnwidth]{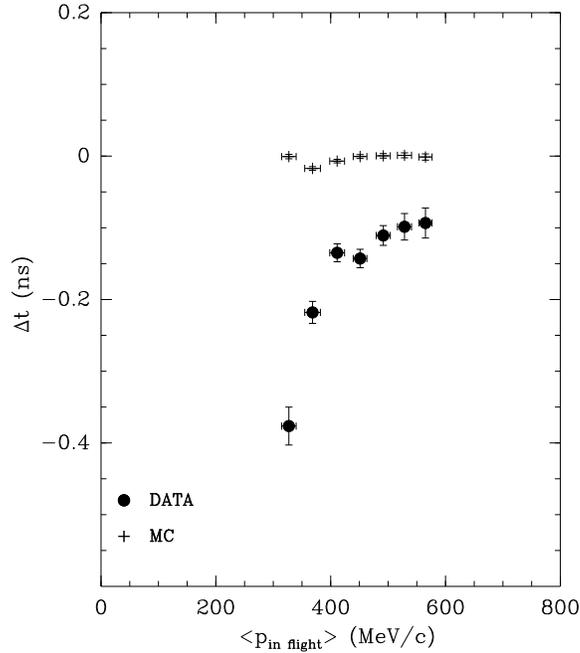}
\caption {  The difference of the time offset measured
 in pad ring 3 from the  expected time offset for protons as a
 function of the 
momentum along its 
 flight path  (in the gas volume of the TPC). The filled circles
 show the results of measurements using elastic scattering on hydrogen, the
 points without marker represent the simulation of the  measurement using the same
 reconstruction procedure. 
 The momentum was predicted using the kinematics of elastic scattering.
 Consistency of the simulated time difference with zero shows
 that the prediction of the flight time (and thus of the momenta) using
 the elastic scattering kinematics and Monte Carlo
 corrections in the reconstruction procedure for respective energy
 losses are correct. From reference~\cite{phys-RPC}.
}
\label{fig:elas:dt:diff}
\end{figure}

Results of this analysis are shown in Fig.~\ref{fig:elas:dt:diff}.
Due to the kinematics of elastic scattering the vast majority of
selected recoil protons which reach the RPC system is measured in pad
ring 3.

The data exhibit a clear deviation pointing
to a difference in RPC time response to protons as a function of the
momentum.
The difference can only be due to the different response of the RPCs to
heavily ionizing compared to minimum ionizing  particles.
The observed effect accounts for the largest fraction of the absolute
values and the shape of the deviations observed in
Fig.~\ref{fig:RPC}. 

The remaining difference observed between the points of
Fig.~\ref{fig:RPC} and Fig.~\ref{fig:elas:dt:diff} is of the order of
($150 \pm 100$)~ps at 450~\MeVc, where the error is estimated from the
spread of the points for the different pad rings.
The central value of 150~ps corresponds to a momentum shift of
$\approx$4.5\% at 450~\MeVc. 

Several important systematic errors affect this measurement:
\begin{itemize}
 \item
      The momentum prediction with elastic scattering needs a correction for
      energy loss in the region of the inner field cage of the TPC.  
      Although the description of the physical processes is very
      accurate it is possible that a 
      slightly larger amount of
      material is present
      than that accounted for in the calculations (the opposite is
      excluded). 
      If the calculation is repeated with 10\% more material a 1\%--2\%
      shift in predicted momentum is induced which would {\em reduce} the
      apparent difference.
 \item 
      Background hits in the RPC pads can only create an earlier time
      measurement, since single-hit TDCs were used to read out the
      system. 
      Given the charged track multiplicities and the
      corresponding number of converted photons from \piz decays this
      overlap probability is estimated to be $\approx 5$\%.
      The effect of such background is not easy to estimate, but the
      resulting measurement is shifted towards shorter time-of-arrival. 
      This background is not present in elastic scattering events.
\item
    There is a 20 ps--30 ps difference in measured arrival time for
    $\sim 400$ \MeVc \pip versus \pim.
    However in the negative beams this is as small as 0 ps--10 ps.
   The difference with opposite B field shows already that the RPCs have
    this kind of systematic.
    The latter can be due to the position-dependent slewing correction
     to the amplifier position which has as maximum swing 180 ps,
     and assumes exact knowledge of where the first electron was detected.
    (there is a symmetry breaking due to the amplifier position
    always to one side.)
\item
   There is a  1\% difference in the average pulse-height for \pip and \pim
   with $\sim 400$ \MeVc in the 90$^\circ$ direction.
   There the production cross-sections are equal.
   The difference can come from the $E\times B$  effect for the avalanche electrons
   which can induce a different space-charge effect due to the different angle
     of incidence of the \pip and \pim in the RPC measurement gap due to
     the opposite curvature of their trajectories in the TPC.  
   This can explain a $\pm 15$ ps difference of threshold crossing,
     keeping in mind that the threshold was relatively high,
   considering that the full time slewing correction is $\sim 2000$ ps.
\end{itemize}

Therefore, the RPC system cannot provide a cross-check of the same quality
as the elastic scattering data.
As a conclusion, the observed time-of-arrival of protons at the RPC
barrel is consistent with an unbiased measurement of momentum within a
relatively large error of 5\%.


\section{Conclusions}
\label{section:conclusion}
Asserting the correctness of the momentum reconstruction in the HARP TPC 
has not been easy, as can be expected from a chamber affected by a large
 number of dead channels, cross-talk, static and dynamic distortions in
 the absence of the possibility to use a direct particle beam for
 calibration. 
 By a series of dedicated cross-checks with benchmarks, the experimental
 verification could nevertheless be made. This allowed us to conclude
 that the TPC momentum reconstruction developed 
 by the HARP collaboration is correct within the precision of $\pm$3\%. 
 This confirms the systematic error associated to the momentum scale used
 in determining  the large angle
  production of charged pions by protons in
  \cite{Ta-paper, ref:HARP:carbon}.

The calibrations and cross-checks include
\begin{itemize}
	\item
reconstruction of the missing mass squared of pp elastic scattering
data;
	\item
comparison of the momentum of the proton scattered at large angle
as measured by the TPC and as calculated from the scattering
angle of the forward particle in pp and $\pi^{\pm}$p elastic scattering 
events;
	\item
dependence of residuals upon polar angle and upon magnetic field
polarity reversal, for tracks reconstructed with and without vertex
constraint during the fit;
	\item
absence of slope  in the momentum versus $\sin \theta$ plots in a fixed
slice of \dedx; 
	\item
comparison of the \dedx curves in the region of high ionization where the
 ionization varies very quickly with momentum, allowing a sensitive
 verification of the momentum scale.
\end{itemize}

We also revisited methods of lesser precision, such as \dedx in the region
 near the minimum ionization, for which we found that it is crucial to use
 a complete Bethe-Bloch formula to 
reach reasonable conclusions.
Once this is done we find 
a good match between \dedx theoretical curves and our data, in comfort of our
 momentum reconstruction.

Finally a careful analysis of the time response of the RPC system
ascertains that no momentum bias is present beyond the uncertainties of
this method.
While investigating any possibility of systematic effect on the momentum
measurement, the presence of a systematic effect in the time measurement
of the RPCs has been demonstrated.


As a conclusion, none of the benchmarks has revealed any significant
bias in the momentum measurement beyond a systematic error of 3\% for
the momentum scale in the TPC.


\renewcommand{\theequation}{A-\arabic{equation}}
  \setcounter{equation}{0}  

\section{Acknowledgments}
%
We gratefully acknowledge the help and support of the PS beam staff
and of the numerous technical collaborators who contributed to the
detector design, construction, commissioning and operation.  
 
In particular, we would like to thank
G.~Barichello,
R.~Brocard,
K.~Burin,
V.~Carassiti,
F.~Chignoli,
D.~Conventi,
G.~Decreuse,
M.~Delattre,
C.~Detraz,  
A.~Domeniconi,
M.~Dwuznik,   
F.~Evangelisti,
B.~Friend,
A.~Iaciofano,
I.~Krasin, 
D.~Lacroix,
J.-C.~Legrand,
M.~Lobello, 
M.~Lollo,
J.~Loquet,
F.~Marinilli,
J.~Mulon,
L.~Musa,
R.~Nicholson,
A.~Pepato,
P.~Petev, 
X.~Pons,
I.~Rusinov,
M.~Scandurra,
E.~Usenko,
and
R.~van der Vlugt,
for their support in the construction of the detector.
 
The collaboration acknowledges the major contributions and advice of
M.~Baldo-Ceolin, 
M.T.~Muciaccia and A. Pullia
during the construction of the experiment.
 
The collaboration is indebted to 
V.~Ableev,
P.~Arce,   
F.~Bergsma,
P.~Binko,
E.~Boter,
C.~Buttar,  
M.~Calvi, 
M.~Campanelli, 
C.~Cavion, 
A.~Chukanov, 
A.~De~Min,    
M.~Doucet,
D.~D\"{u}llmann,
R.~Engel,   
V.~Ermilova, 
W.~Flegel,
P.~Gruber,   
Y.~Hayato,
P.~Hodgson,  
A.~Ichikawa,
A.~Ivanchenko,
I.~Kato,  
O.~Klimov,
T.~Kobayashi,
D.~Kustov,
M.~Laveder,  
L.~Linssen,
M.~Mass,
H.~Meinhard,
T.~Nakaya,
K.~Nishikawa,
M.~Paganoni,     
F.~Paleari,  
M.~Pasquali,
J.~Pasternak,   
C.~Pattison,    
M.~Placentino,
S.~Robbins,   
G.~Santin,  
S.~Simone,
A.~Tornero,   
S.~Troquereau,
S.~Ueda, 
A.~Valassi,
F.~Vannucci   
and
K.~Zuber   
for their contributions to the experiment and to P. Dini for his
contribution to MC production.

We acknowledge the contributions of 
V.~Ammosov,
G.~Chelkov,
D.~Dedovich,
F.~Dydak,
M.~Gostkin,
A.~Guskov, 
D.~Khartchenko, 
V.~Koreshev,
Z.~Kroumchtein,
I.~Nefedov,
A.~Semak, 
J.~Wotschack,
V.~Zaets and
A.~Zhemchugov
to the construction and operation of the HARP detector.

 The experiment was made possible by grants from
the Institut Interuniversitaire des Sciences Nucl\'eair\-es and the
Interuniversitair Instituut voor Kernwetenschappen (Belgium), 
Ministerio de Educacion y Ciencia, Grant FPA2003-06921-c02-02 and
Generalitat Valenciana, grant GV00-054-1,
CERN (Geneva, Switzerland), 
the German Bundesministerium f\"ur Bildung und Forschung (Germany), 
the Istituto Na\-zio\-na\-le di Fisica Nucleare (Italy), 
INR RAS (Moscow) and the Particle Physics and Astronomy Research Council (UK).
We gratefully acknowledge their support.
This work was supported in part by the Swiss National Science Foundation
and the Swiss Agency for Development and Cooperation in the framework of
the programme SCOPES - Scientific co-operation between Eastern Europe
and Switzerland. 
\appendix
\section{Appendix: Treatment of the Dynamic Distortions \label{Appendix}}
Given the beam intensity, the data acquisition rate and the target
length (5\% of the nuclear interaction length), it is computed that HARP
operated dead time larger than 90\%. 
The electrons are normally amplified near the TPC pad plane with an amplification factor of the order of $10^5$, producing an equivalent number of Argon ions. 
Any inefficiency of the gating grid at the level of $10^{-3}$ or even $10^{-4}$
would let an overwhelming number of ions drift into the TPC gas volume.

\begin{figure}
\begin{center}
\includegraphics[width=0.43\textwidth,angle=0]{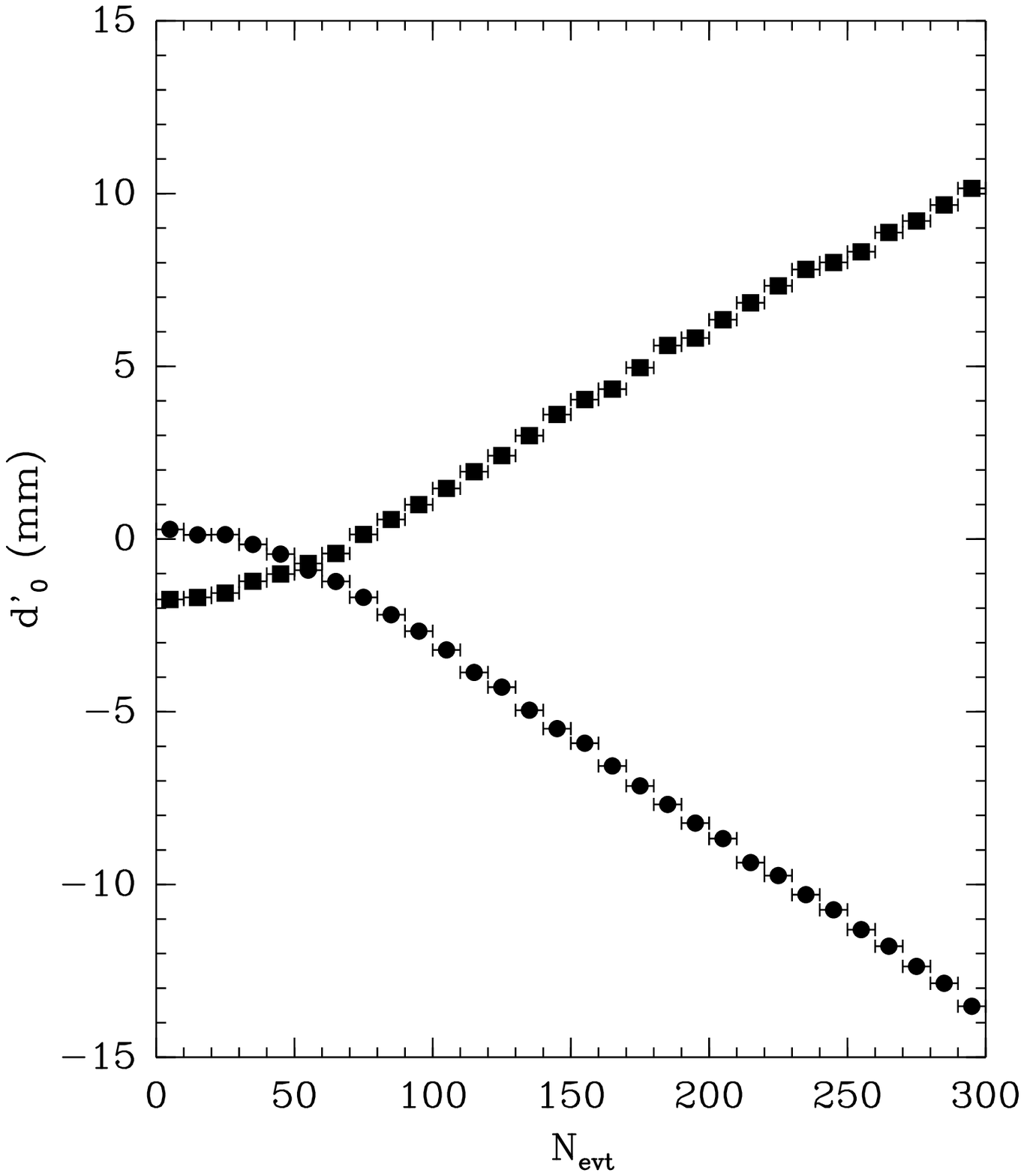}
\includegraphics[width=0.43\textwidth,angle=0]{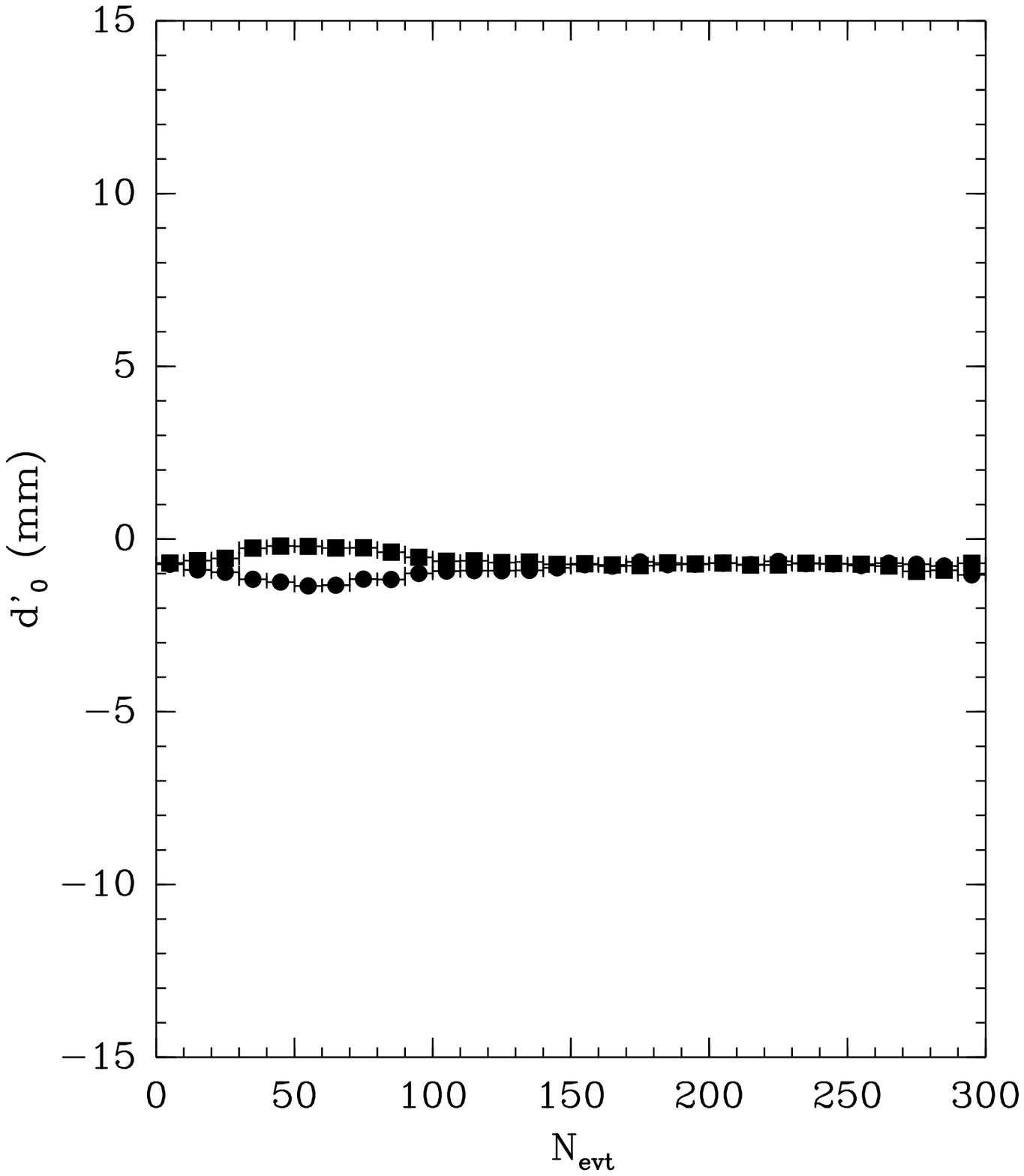}
\end{center}
\caption{
   Average \dzeroprime as a function of event number in spill for 8.9~\GeVc
   Be data. (left panel uncorrected; right panel: dynamic distortion
   corrections applied.)
   After the ``default'' correction for the static distortions (equal for
   each setting) a small residual effect at the beginning of the spill is
   visible at $\evtspill=0$ (left panel).
   This is due to the fact that the inner and outer field cages were powered
   with individual HV supplies.
   A setting-by-setting correction compatible with the reproducibility of
   the power supplies is applied for the data of the right panel together
   with the dynamic distortion correction.
}
\label{fig:be:dzeroprime}
\end{figure}

This indeed turns out to be the case.
The dynamic distortions
can be monitored using the average value of the extrapolated minimum
distance of secondary tracks from the incoming beam particle trajectory
 \dzeroprime.
This is a similar procedure as the one being used for the STAR
TPC~\cite{ref:star}. 
Using calibration data sets, the deterioration of the performance of the
detector, see Fig.~\ref{fig:be:dzeroprime} (left), is determined
as a function of the strength of the distortions characterized by an average
value of \dzeroprime: for each particular setting only that
part of the data for which the systematic error was under control 
was used for the first
analysis (of the order of 30\% of available statistics)~\cite{Ta-paper,ref:HARP:carbon}.
As a second step, a physics model fully describing the time development
 of dynamic distortions during physics spills has been developed and
 benchmarked, as well as a correction
 algorithm~\cite{Dynamic-Distortion} implemented. 

\begin{figure}[tb]
\centering
\includegraphics[width=0.40\columnwidth]{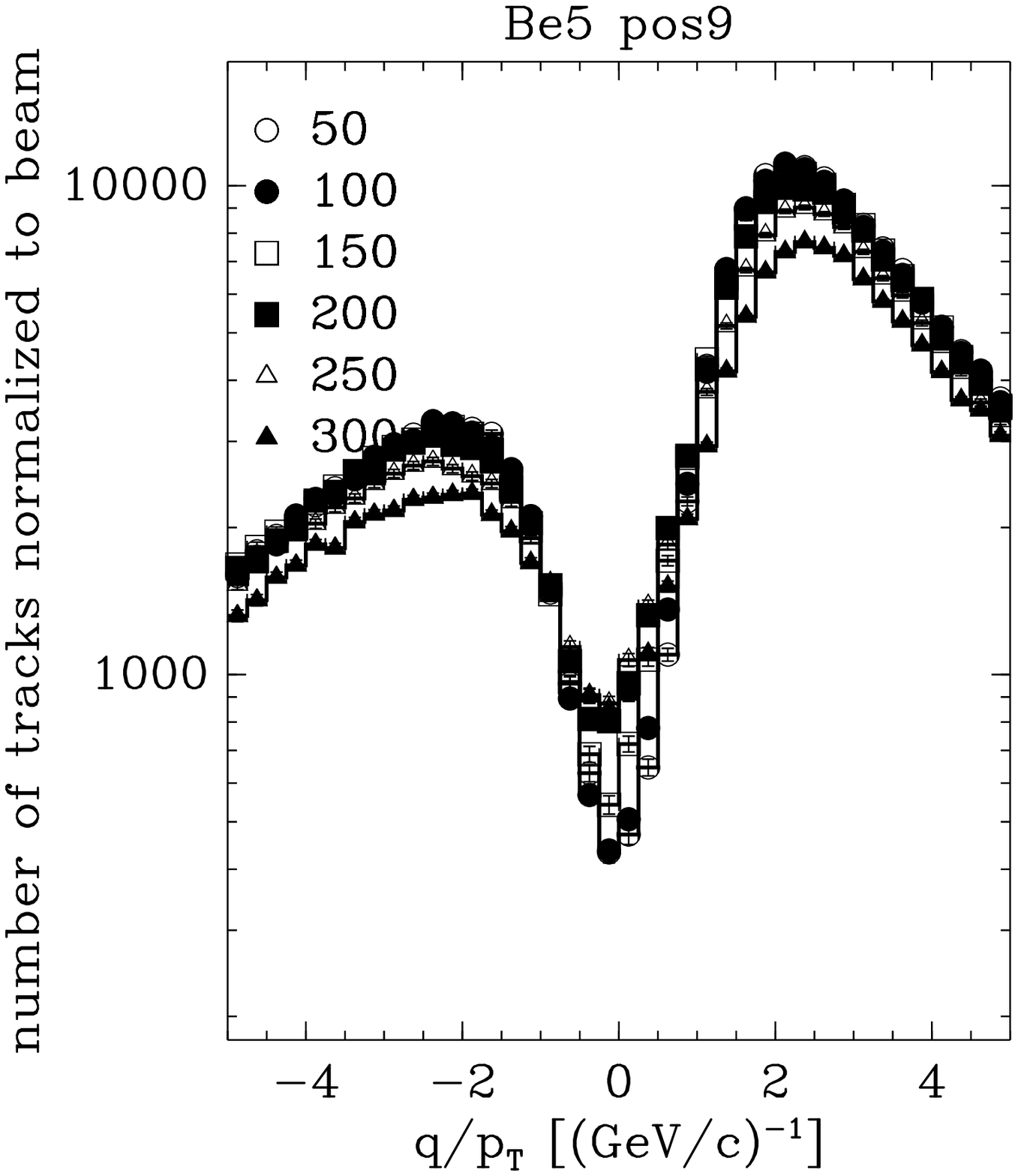}
\includegraphics[width=0.40\columnwidth]{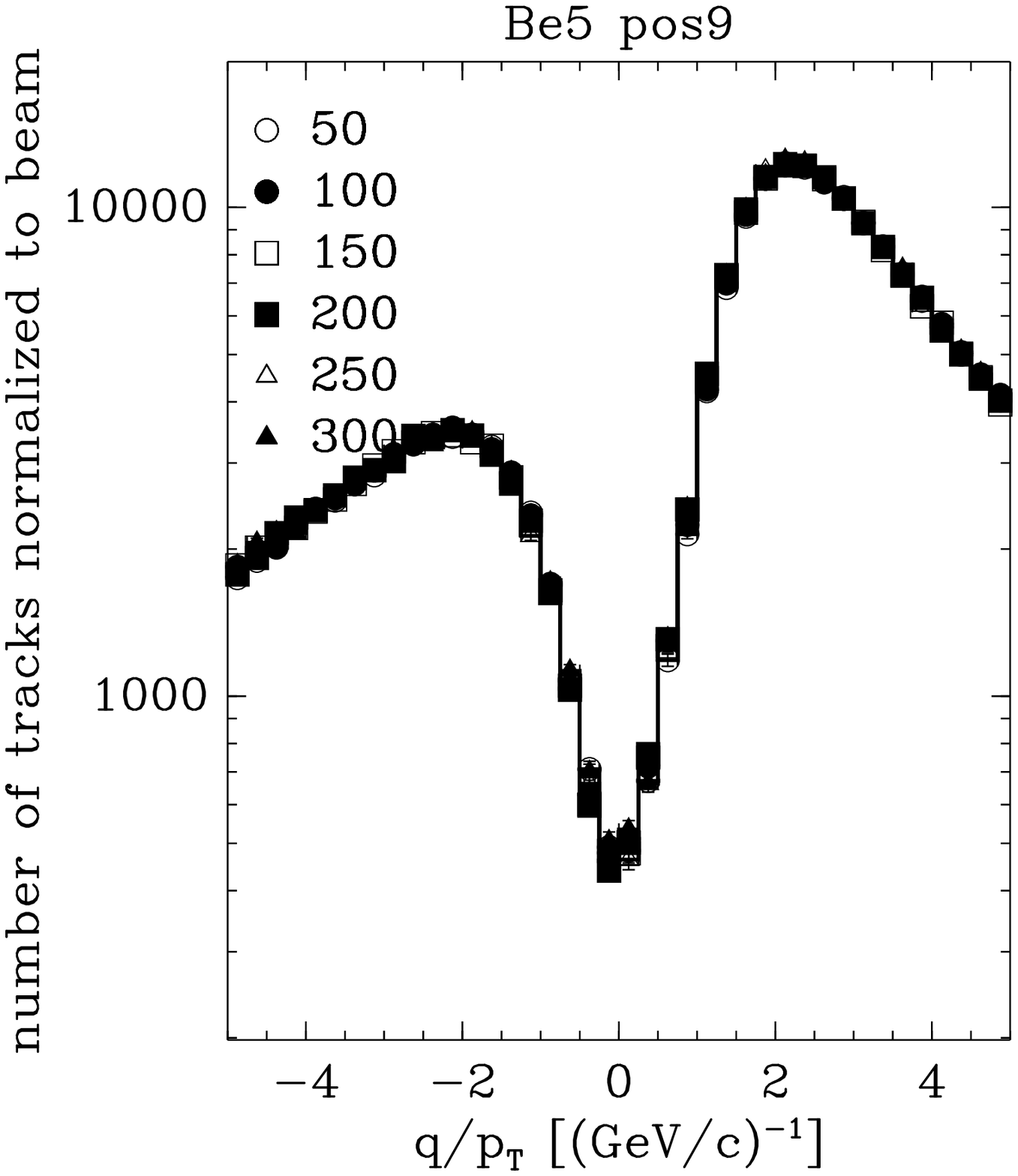}
\caption{
  Analysis of $Q/\pt$ for the highest statistics
  data sample: p-Be at 8.9~\GeVc. Left panel: distortions are not corrected;
  six curves are drawn, each for the next 50 events in the spill.
  Right panel: dynamical distortions are corrected; the six curves are almost
  not distinguishable  }
\label{fig:full-spill-calib}
\end{figure}

In addition to the physics model, direct measurements of the
displacements of the positions measured at the pad plane of the TPC were
performed by predicting the full track trajectory in space using elastic
scattering kinematics.
The direct measurement and the model show good agreement, indicating
that the effect is fully understood.
The effects of this correction can be appreciated in
 Fig.~\ref{fig:be:dzeroprime} (right).
The comparison of results obtained using the uncorrected first part of
 the spill, as in the first HARP analysis, with those using the full
 corrected spill (see  Fig.~\ref{fig:full-spill-calib})
 shows excellent agreement. This provides an {\sl a posteriori}
 confirmation with 2 to 3 times better statistics
 that the approach used in the first HARP analysis was correct.
This is not unexpected, since, owing to their limited mobility the first ions
created in the amplification region need about 25 ms to reach the drift
region and subsequently the steady flow of ions into this region only
starts approximately 100 ms after the start of the spill, with a gradual
transition between these two regimes.

\end{document}